\documentclass[%
reprint,
superscriptaddress,
amsmath,
amssymb,
aps,
]{revtex4-2}
\usepackage[utf8]{inputenc}
\usepackage[T1]{fontenc}
\usepackage{color}
\usepackage{graphicx}
\usepackage{dcolumn}
\usepackage{bm}
\usepackage{xfrac}
\usepackage{upgreek}

\usepackage{dcolumn}
\usepackage{bm}
\usepackage{braket}
\usepackage{sidecap}
\usepackage {hyperref}
\usepackage {notes2bib}

\graphicspath{Figures}

\begin{document}

\title{Analysis of the exciton fine-structure in InAs quantum dots with cavity-enhanced emission at telecom wavelength and grown on a GaAs(111)$A$ vicinal substrate}

\author{
A. Barbiero$^{1,2 \dagger\star}$,
A. Tuktamyshev$^{3,4 \dagger\ast}$,
G. Pirard$^{5}$,
J. Huwer$^{1}$,
T. M\"{u}ller$^{1}$,
R. M. Stevenson$^{1}$,
S. Bietti$^{3},$
S. Vichi$^{3,4},$
A. Fedorov$^{6},$
G. Bester$^{5},$
S. Sanguinetti$^{3,6}$
and A. J. Shields
}

\affiliation{\small
Toshiba Europe Limited, 208 Cambridge Science Park, Cambridge CB4 0GZ, UK \\
$^{2}$Department of Physics and Astronomy, University of Sheffield, Hounsfield Road, Sheffield S3 7RH, UK \\
$^{3}$L–NESS and Material Science Department, Universita` degli Studi di Milano Bicocca, via R. Cozzi 55, 20125 Milano, Italy \\
$^{4}$INFN - Sezione di Milano Bicocca, Piazza della Scienza 3, Milano, Italy \\
$^{5}$Physical Chemistry and Physics Departments, University of Hamburg, Luruper Chaussee 149, Hamburg, Germany \\
$^{6}$L-NESS and CNR-IFN, Piazza Leonardo da Vinci 32, Milano, Italy \\
\medskip
$^{\dagger}$These authors contributed equally to this work;\\
$^{\star}$andrea.barbiero@crl.toshiba.co.uk;
$^{\ast}$artur.tuktamyshev@unimib.it\\
}
\medskip

\begin{abstract}
\noindent The efficient generation of entangled photons at telecom wavelength is crucial for the success of many quantum communication protocols and the development of fiber-based quantum networks. Entangled light can be generated by solid state quantum emitters with naturally low fine-structure splitting, such as highly symmetric InAs quantum dots (QDs) grown on (111)-oriented surfaces. Incorporating those QDs into optical cavities is crucial to achieve sufficient signal intensities for applications, but has so far shown major complications. In this work we present droplet epitaxy of telecom wavelength InAs QDs within an optical cavity on a vicinal (2\textdegree{} miscut) GaAs(111)$A$ substrate. We show a remarkable enhancement of the photon extraction efficiency compared to previous reports together with a reduction of the density that facilitates the isolation of single spectral lines. Moreover, we characterize the exciton fine structure splitting and employ numerical simulations under the framework of the empirical pseudopotential and configuration interaction methods to study the impact of the miscut on the optical properties of the QDs. We demonstrate that the presence of the miscut steps influences the polarization of the neutral excitons and introduces a preferential orientation in the C\textsubscript{3v} symmetry of the surface.
\end{abstract}

\maketitle


\section{Introduction}
\noindent Single and entangled photon emitters are fundamental building blocks for emerging technologies such as quantum communication protocols and quantum networks \cite{Kimble.2008,Wehner.2018}. In this framework, sources of entangled photon pairs based on the biexciton (XX) – exciton (X) recombination cascade in semiconductor quantum dots (QDs) \cite{SkibaSzymanska.2017,Orieux.2017,Huber.2018} offer multiple advantages such as electrical control, tunability and integration with various photonic structures \cite{Kirsanske.2017,Uppu.2020,Thomas.2021b,Tomm.2021,Sartison.2017,Chen.2018,Kaganskiy.2018}. However, QDs typically show a fine-structure splitting (FSS) between the exciton eigenstates larger than the linewidth of the transition \cite{Gammon.1996, Kulakovskii.1999}, which is caused by anisotropies in their shape or composition and complicates entanglement-based experiments. One solution for the development of highly symmetric QDs with low FSS is to self-assemble them on (111)-oriented substrates due to the natural C\textsubscript{3v} symmetry of the surface \cite{Mano.2010,Kuroda.2013,BassoBasset.2018}.

Moreover, for long-distance quantum communication and integration with the existing fiber infrastructure it is necessary to select a platform that provides emission at telecom wavelength, such as InAs/GaAs QDs \cite{Ha.2015}. It is worth noting that InAs QDs cannot be grown on GaAs(111) with the common Stranski-Krastanov (SK) method. In fact, InAs heteroepitaxy on GaAs(111)$A$ proceeds with the formation of misfit dislocation at the InAs/GaAs interface, resulting in an effective strain relaxation without nucleation of 3D islands \cite{Yamaguchi.1997}. Instead, they require the more advanced Droplet Epitaxy (DE) technique \cite{Koguchi.1991, Gurioli.2019}, which relies on the formation of group-III metal droplets followed by crystallization in a group-V atmosphere. DE offers better control over the QD self-assembly dynamics and guarantees fine tuning of the shape, size and density of the nanostructures \cite{Somaschini.2009,Somaschini.2010}.
\begin{figure*}
\includegraphics[width=0.67\textwidth]{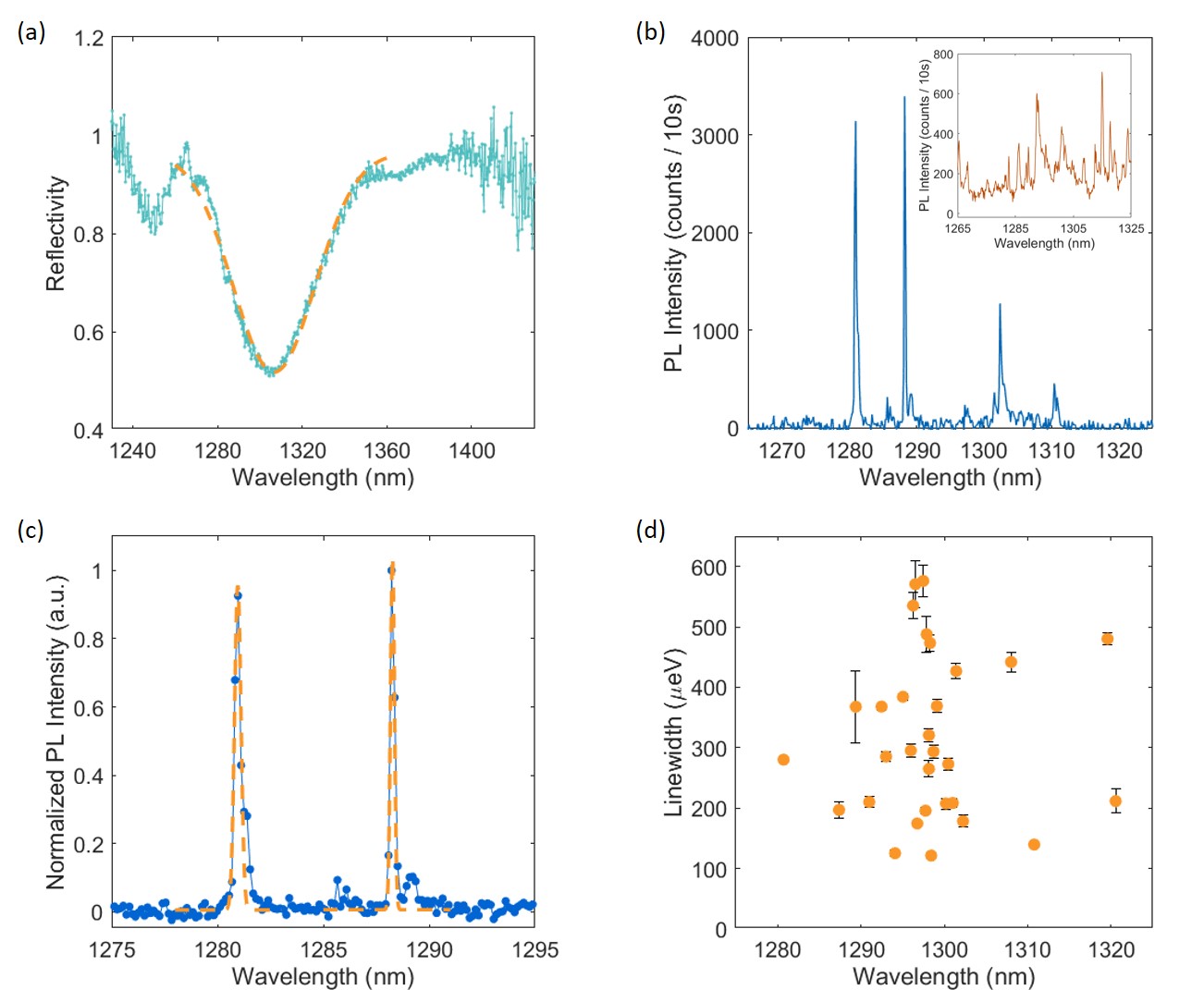}
\caption{(a) Reflectivity measurement taken on the optical cavity and corresponding Gaussian fit (dashed line), showing a mode centered at 1307 nm with FWHM = 39 nm. (b) Typical PL spectrum of an individual InAs QD within the optical cavity under CW laser excitation at 785 nm. The inset shows an example of PL spectrum measured in equivalent conditions on a different sample \cite{Tuktamyshev.2021} with the same InAs QDs but no DBR mirrors. (c) Gaussian fit of two transitions from the spectrum reported in (b). (d) Linewidth statistics obtained by fitting 30 spectral lines.
\label{Fig1} }
\end{figure*}
The formation of telecom wavelength InAs QDs on GaAs(111)$A$ with FSS as low as 16 $\mu$eV has been recently demonstrated \cite{Tuktamyshev.2021}. Nevertheless, previous works on (100) substrates suggest that incorporating those emitters in a one-dimensional microcavity could improve the photon extraction efficiency and generate sufficient signal intensities for applications \cite{Ward.2014, Huwer.2017, Muller.2018}. Unfortunately, the deposition of distributed Bragg reflectors (DBRs) on singular, i.e. on-axis, (111) substrates has proven challenging because very low growth rates below 0.03 nm/s are required to obtain flat epilayers \cite{Esposito.2017}.

Here, we demonstrate self-assembly of InAs DE QDs emitting in the telecom O-band within a DBR microcavity grown on a vicinal (2\textdegree{} miscut) GaAs(111)A substrate. Thanks to the presence of the miscut, the growth rates can be increased by up to one order of magnitude (0.14 – 0.28 nm/s) and therefore become similar to the ones used on standard GaAs(100) \cite{Herzog.2012,Tuktamyshev.2019}. Moreover the propensity of twin defects in the AlGaAs layers, which often limits the quality of structures grown on singular (111) substrates \cite{Herzog.2012}, is strongly reduced.
Combining the DE technique with the growth of a DBR microcavity enables us to meet the high-brightness and low-density criteria necessary for the spectroscopic investigation of single QDs.
%

\section{Epitaxy}
\noindent The sample was grown on an undoped vicinal GaAs(111)$A$ substrate with a 2\textdegree{} miscut along the [$\bar{1}\bar{1}2$] direction in a solid source Molecular Beam Epitaxy (MBE) system.
After a GaAs buffer, a bottom DBR with 25 repeats of $\lambda/4$ Al\textsubscript{0.5}Ga\textsubscript{0.5}As/GaAs layers was grown at 600\textdegree C using deposition rates of 0.28 nm/s and 0.14 nm/s respectively. Such a DBR provides high reflectivity for a wavelength range centered around 1310 nm.
Next, DE InAs QDs with an approximate density of 1×10\textsuperscript{8} cm\textsuperscript{-2} were formed in the middle of a $\lambda/2$ cavity consisting of 201.6 nm of In\textsubscript{0.6}Al\textsubscript{0.4}As (grown at 470\textdegree C with a deposition rate of 0.14 nm/s) by supplying 1 ML (monolayer) of In at 370\textdegree C with a deposition rate of 0.003 nm/s and then crystallizing for 8 minutes in As\textsubscript{4} atmosphere at 300\textdegree C. During the In deposition the residual As\textsubscript{4} beam equivalent pressure (BEP) was kept below 3×10\textsuperscript{-9} Torr. The As\textsubscript{4} valve was then opened for crystallization, resulting in a BEP of 3×10\textsuperscript{-5} Torr. These parameters are similar to those previously developed for growth on singular GaAs(111) substrates \cite{Ha.2015}.
Finally, three $\lambda/4$ layers (GaAs/Al\textsubscript{0.5}Ga\textsubscript{0.5}As/GaAs) were deposited with the same conditions used for the bottom DBR \bibnote{See Supplemental Material at [URL will be inserted by publisher] for further details on the sample growth and morphology.}. The asymmetric cavity was designed to limit the photon leakage into the substrate and direct the emission towards the collection optics.

\section{Optical characterization}
\begin{figure*}
\includegraphics[width=0.67\textwidth]{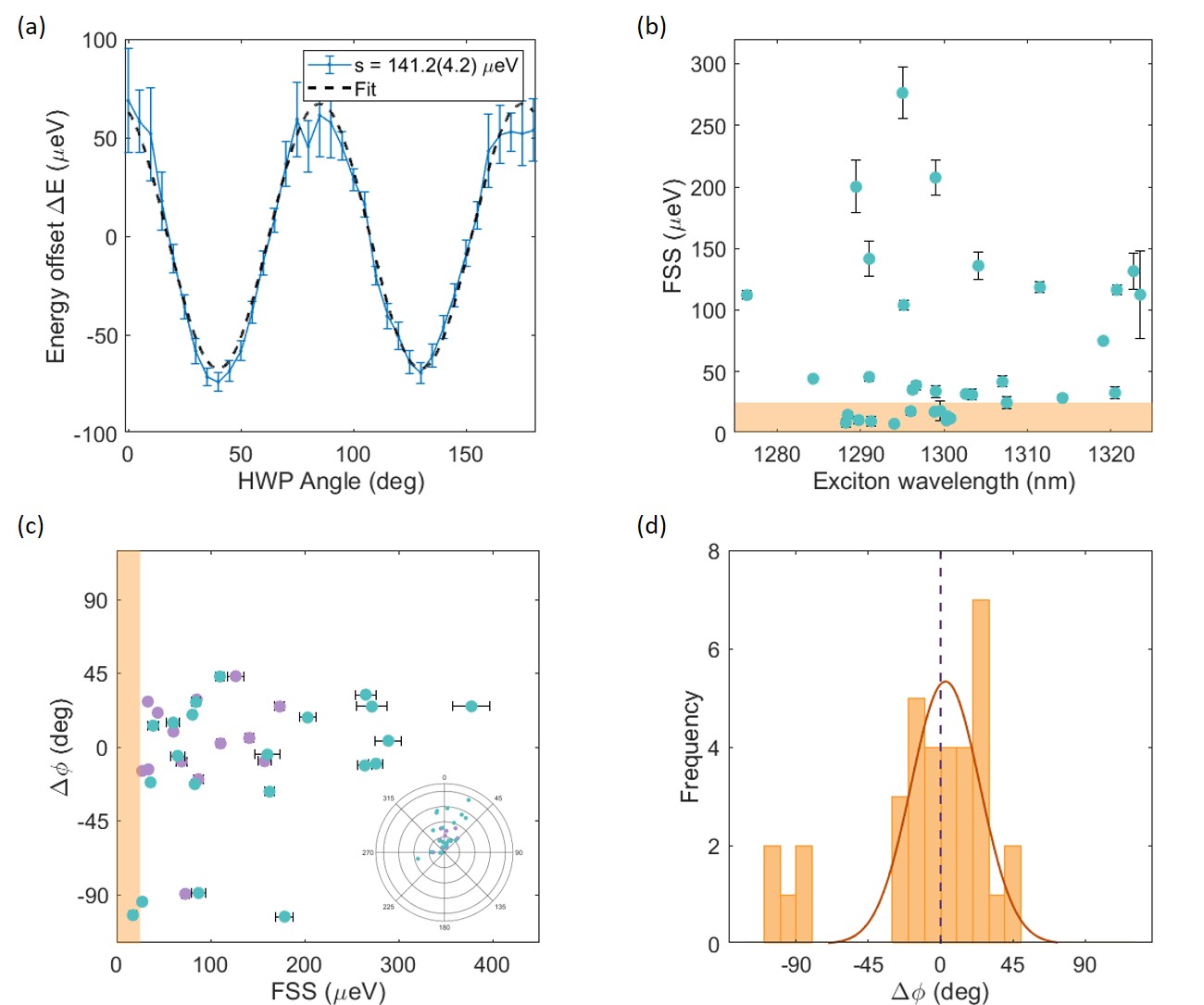}
\caption{(a) Example of FSS measured on an InAs QD emitting in the telecom O-band using the HWP method. The dashed line is the sinusoidal fit of the energy offset. (b) Statistical distribution obtained by measuring the FSS of 35 spectral lines of unidentified species. The orange area indicates the estimated resolution limit for a transition with 250 $\mu$eV linewidth. (c) Dipole orientation of the neutral excitions as a function of the FSS measured on 35 QDs. $\Delta\phi$ is the angular offset with respect to the [$\bar{1}$$\bar{1}$2] direction while the orange area indicates the estimated resolution limit for a transition with 250 $\mu$eV linewidth. The data points are plotted in violet if it was possible to identify clearly a XX-X pair and in blue if FSS and $\Delta\phi$ were extracted from one line only. (d) Corresponding histogram with Gaussian fit of the statistical distribution. The vertical dashed line shows the direction of the miscut steps.
\label{Fig2} }
\end{figure*}
\noindent For optical characterization, the sample was mounted inside a closed-cycle cryostat at a temperature of 5 K and the QDs were excited by a CW laser at 785 nm. The emitted photons were collected using a fiber-coupled confocal microscope and sent to a spectrometer equipped with an InGaAs photodiode array for analysis.

Figure \ref{Fig1}a shows a wide-range reflectivity measurement taken on the sample. The dip in the spectrum confirms the presence of a cavity mode at the desired wavelength, which is fitted with a Gaussian curve centered at 1307 nm. The extracted FWHM = 39 nm is similar to the value predicted by our simulations.

In Figure \ref{Fig1}b we report a typical O-band photoluminescence (PL) spectrum from the QDs. The inset shows an example of PL measurement taken in the same setup on InAs/InAlAs QDs grown without DBR mirrors \cite{Tuktamyshev.2021}, confirming that the presence of the optical cavity leads to an enhancement of the signal intensity by a factor > 5. We also notice an improvement in the signal/background ratio and an important reduction in the density of spectral lines, which makes it possible to isolate the emission pattern of single QDs.
The FWHM of the spectral lines is extracted with a Gaussian fit, as shown in Figure \ref{Fig1}c. Fitting 30 different transitions reveals no correlation between the emission wavelength and the linewidths, which vary between 100 $\mu$eV and 550 $\mu$eV (Figure \ref{Fig1}d). Those rather broad linewidths are consistent with previously reported values \cite{Tuktamyshev.2021} and originate from the presence of point defects and threading dislocations in the InAlAs barrier layers.

To quantify the FSS we employed the well-established half-wave plate (HWP) method \cite{Young.2005,SkibaSzymanska.2017}: we sent the PL signal through a rotating HWP and a fixed linear polarizer (LP) mounted in the collection arm of the confocal microscope and measured the resulting shift in energy at the spectrometer. An example of FSS measurement is shown in Figure \ref{Fig2}a, where the energy shift as a function of the HWP angle fits well with the expected sinusoidal behavior.
A statistical distribution was obtained by repeating the measurement on 35 lines of unidentified species in the telecom O-band: as reported in Figure \ref{Fig2}b, approximately 50\% of those lines show a FSS < 50 $\mu$eV, while larger values between 70 $\mu$eV and 300 $\mu$eV are recorded in the other cases. In general, these values are similar to the ones reported in literature for DE QDs grown on GaAs(100) substrates \cite{SkibaSzymanska.2017}.
\begin{figure*}
\includegraphics[width=0.75\textwidth]{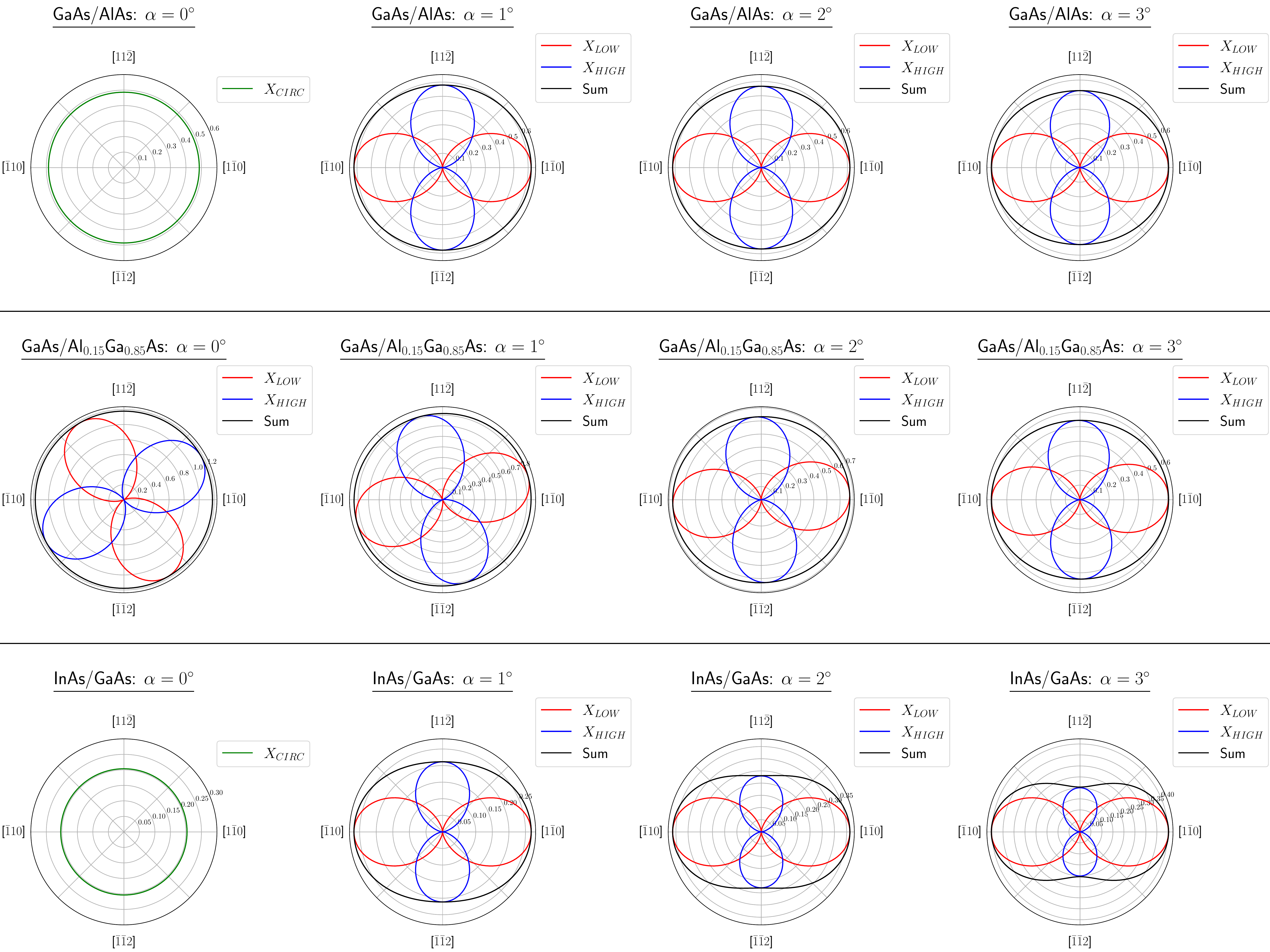}
\caption{Polar diagrams showing the square of the optical transition dipole matrix element calculated for GaAs/AlAs (top panel), GaAs/Al\textsubscript{0.15}Ga\textsubscript{0.85}As (middle panel) and InAs/GaAs (bottom panel) QDs assuming different values of the miscut angle $\alpha$. 'X\textsubscript{LOW}' ('X\textsubscript{HIGH}') represents the lowest (highest) lying bright exciton state in energy, whereas two degenerate exciton bright states are labelled 'X\textsubscript{CIRC}'. The black line labeled 'Sum' represents the net total polarization from which the DLP is extracted.
\label{Fig3} }
\end{figure*}

The presence of emitters with large FSS may originate from unexpected anisotropy in some of the QDs examined. To gain more insight into this phenomenon, we carried out an additional investigation focusing on the orientation of the neutral excitons. First, we adjusted the position of the sample in the cryostat such that it was possible to identify the direction of the miscut steps with respect to the lab frame and to the optics mounted in the collection arm of the confocal microscope.
Then we repeated the FSS measurements over different QDs and selected 35 neutrally charged excitons with splitting larger than 10\% of the linewidth.
For the majority of those emitters we found that the oscillations of the energy offset display the first maximum for a similar HWP rotation, suggesting that the dipoles are aligned along one preferential direction (Figure \ref{Fig2}c). No correlation between the FSS magnitude and the dipole orientation was observed.

In Figure \ref{Fig2}d we report a histogram of the dipole orientations: the Gaussian fit shows that the statistical distribution is centered around $\Delta\phi$ = 3.1\textdegree ($\pm$ 2.2\textdegree) indicating that, for the majority of the QDs, the component of the X doublet higher in energy tends to be polarized along [$11\bar{2}$]. In fact, the small discrepancy of $\sim$ 3\textdegree{} can easily be attributed to limited accuracy while mounting the sample.
Interestingly, a small group of QDs exhibits perpendicular polarization aligned with [$1\bar{1}0$]. Since for many emitters it was not possible to identify clearly a XX-X pair and the values of FSS and $\Delta\phi$ were extracted from one spectral line only, it may be possible that those data points originate from a XX transition and consequently have the highest energy component of the doublet oriented at 90\textdegree{} with respect to the X due to conservation of energy and spin.
In summary, these results suggest that the presence of a 2\textdegree{} miscut introduces a privileged direction in the natural C\textsubscript{3v} symmetry of the system.
\section{Influence of the miscut on the optical properties of the QDs}
\noindent Following the experimental observations presented above, we carried out numerical simulations to better understand the impact of the miscut on the optical properties of the QDs \bibnote{See Supplemental Material at [URL will be inserted by publisher] for further details on the numerical methods.}. Since the entire structure of the sample including the InAlAs barrier layers and the DBR mirrors could not be reproduced because of computational limitations, we considered 3 simplified material systems. First we studied GaAs/AlAs(111) QDs, which are a simple case with no strain or alloy. Then we examined GaAs/Al\textsubscript{0.15}Ga\textsubscript{0.85}As(111) QDs to isolate the effects of alloying and finally InAs/GaAs(111) QDs, which are an idealized version of the sample and reveal the impact of strain due to lattice mismatch.

The dots were modelled as hexagonally-based truncated pyramids with 70 nm diameter, 4 nm height and a variable miscut angle 0\textdegree{} $ \leq \alpha \leq$ 3\textdegree{} along [$\bar{1}\bar{1}2$]. They were placed in the center of a cubic simulation box filled with the barrier material and the structure was allowed to relax in order to minimise the strain energy using a generalized valence force field (GVFF) model \cite{Keating.1966,Williamson.2000}.
The single-particle Schrödinger equation was then solved (both for the conduction and valence states) under the framework of the empirical pseudopotential method (EPM) \cite{Bester.2009} using a strained linear combination of bulk Bloch bands (SLCBB) \cite{Wang.1999} as a basis to expand the wave functions. Then the many-body problem was addressed through a configuration interaction (CI) scheme \cite{Franceschetti.1999}, with the correlated exciton wave functions described as a linear combination of singly excited Slater determinants. The Hamiltonian was constructed from the electron-hole Coulomb and exchange integrals, which were calculated from the SLCBB wave functions and screened according to the Resta model \cite{Resta.1977}.

In Figure \ref{Fig3} we show the polarization of the two decay paths from the X state, with polar diagrams of the squared optical transition dipole matrix element calculated for the 3 cases described above as a function of the miscut angle $\alpha$. For GaAs/AlAs(111) QDs, characterized by the perfect C\textsubscript{3v} symmetry, the two decay paths from the X state are indistinguishable in absence of miscut ($\alpha$ = 0\textdegree{}) and can be represented by circularly polarized emission.
When the miscut is introduced ($\alpha$ > 0\textdegree{}), the two decay paths become distinguishable and the polarization of the emitted photons reconfigures such that the bright exciton state lower in energy (X\textsubscript{LOW}) is polarized along the [$1\bar{1}0$] crystal axis whereas the the bright exciton state higher in energy (X\textsubscript{HIGH}) is polarized perpendicularly along [$11\bar{2}$]. The net total polarization aligns with [$1\bar{1}0$] and the degree of linear polarization (DLP) increases slightly with the miscut angle.
\\If the AlAs barrier is replaced with Al\textsubscript{0.15}Ga\textsubscript{0.85}As the degeneracy of the exciton states for $\alpha$ = 0\textdegree{} is broken, presumably because of the inhomogeneous composition. Furthermore, the rotation of the polarization as a function of the miscut angle is less pronounced, such that perfect alignment with the crystallographic axes is achieved for larger values of $\alpha$.
Finally, InAs/GaAs(111) QDs show the same behavior of GaAs/AlAs(111) QDs, but with a steeper increase of the DLP as $\alpha$ grows.
\\These simulations explain the experimental results reported in Figure \ref{Fig2}c and \ref{Fig2}d, confirming that the miscut is directly responsible for the appearance of a preferential orientation in all the material systems considered. We observe that the effects of alloying are weaker than the impact of the miscut, which dominates for $\alpha \geq$ 2\textdegree{} and is enhanced by the addition of built-in strain. These findings also support the hypothesis that the 5 data points in Figure \ref{Fig2}c and Figure \ref{Fig2}d with dipoles oriented along [$1\bar{1}0$] are most likely associated to XX transitions, as explained at the end of the previous section.
\begin{figure}
\includegraphics[width=0.5\textwidth]{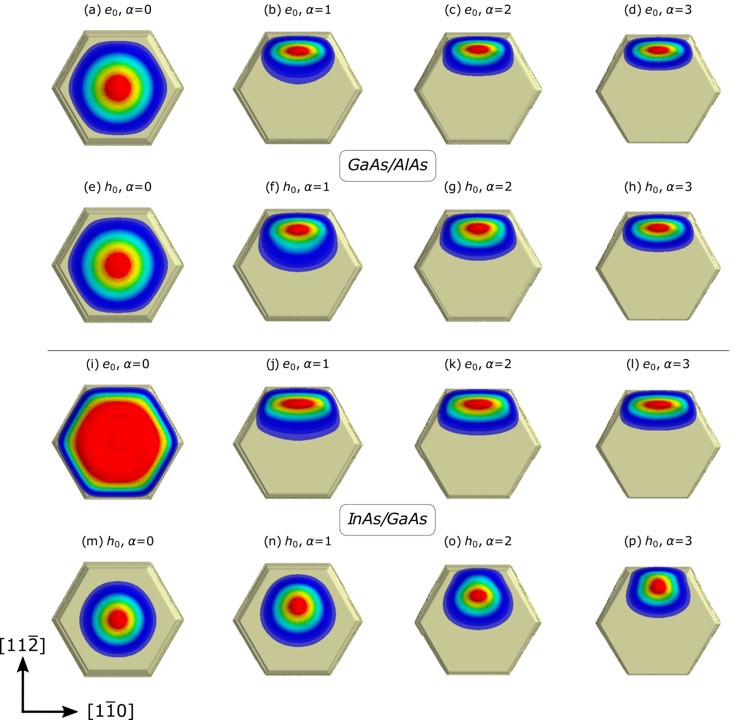}
\caption{Spatial profile of the electron (e\textsubscript{0}) and hole (h\textsubscript{0}) wave functions calculated for GaAs/AlAs QDs (top panel) and InAs/GaAs QDs (bottom panel) with miscut angle $\alpha$ ranging from 0\textdegree to 3\textdegree. The quantisation axis, i.e. [111], is pointing towards the reader.
\label{Fig4} }
\end{figure}

To clarify the physical origin of this behavior, Figure \ref{Fig4} presents the charge densities of the electrons (e\textsubscript{0}) and holes (h\textsubscript{0}) in the conduction and valence ground states for GaAs/AlAs(111) and InAs/GaAs(111) QDs. In strain-free QDs both the electron (Figure \ref{Fig4}a) and hole (Figure \ref{Fig4}e) wave functions exhibit a circular s-like symmetry for $\alpha$ = 0\textdegree. The introduction of the miscut makes the side of the QDs towards [$11\bar{2}$] thicker than the one towards [$\bar{1}\bar{1}2$]. As a result, the wave functions localize at the thicker edge of the dots as $\alpha$ grows and their spatial distribution becomes strongly elongated in the [$1\bar{1}0$] direction. A similar progressive migration towards [$11\bar{2}$] has been observed in the alloyed case, even though the wave functions are not characterized by a circular s-like symmetry.
The picture becomes more complicated in InAs/GaAs(111) QDs: in fact, the electron wave functions can spread more than in the strain-free case and are also found to be more elongated in the [$1\bar{1}0$] direction for $\alpha$ > 0\textdegree{}. At the same time, the hole wave functions are more confined so that their progressive migration towards the thicker edge of the dots as $\alpha$ grows is less pronounced than in GaAs/AlAs(111) QDs and their spatial spread along [$1\bar{1}0$] is reduced.
The elongation of the electron wave function increases with the miscut angle and is responsible for the growth of the intensity of polarization along [$1\bar{1}0$] (\textit{I\textsuperscript{$[1\bar{1}0]$}}) \cite{Narvaez.2005}, in agreement with previous studies on elongated QDs \cite{Sheng.2006}.

Because of the different ability of the electron and hole wave functions to localize at the thicker edge of the InAs/GaAs(111) QDs we observe that there may be a second source of optical anisotropy, which derives directly from the definition of dipole moment. In fact, when the miscut angle increases, the spatial separation between opposite carriers \textit{d\textsubscript{eh}} remains nearly constant in the [$1\bar{1}0$] direction and diminishes along [$11\bar{2}$], which ultimately lowers the intensity of polarization along [$11\bar{2}$] (\textit{I\textsuperscript{$[11\bar{2}]$}}).
The combination of this effect with the elongation of the wave functions discussed above lies behind the optical anisotropy in QDs grown on (111) vicinal substrates and the resulting polarization of the neutrally charged excitons. It is now clear that in GaAs/AlGaAs QDs the anisotropy originates only from the elongation of the electron and hole wave functions which are localized within the same effective volume of interaction. On the other hand, in InAs/GaAs QDs both effects come into play so that \textit{I\textsuperscript{$[11\bar{2}]$}} is modulated by the magnitude of \textit{d\textsubscript{eh}}. As a result, when a 2\textdegree{} miscut is introduced the polarization aligns with the direction of the steps in both cases, but with a different DLP.

Finally, we note that the simplified model presented in this manuscript is intended to provide an intuitive explanation connecting the shape of the envelope functions to the optical properties of the QDs. A rigorous treatment would require taking into account further elements such as the heavy-hole/light-hole mixing and the Bloch functions, which may affect the polarization anisotropy \cite{Tonin.2012}. 
\section{Conclusions}

\noindent In this work we incorporated DE InAs/InAlAs QDs grown on vicinal GaAs(111)$A$ and emitting in the telecom O-band into an optical microcavity based on GaAs/AlGaAs distributed Bragg reflectors. We reported an enhancement of the photoluminescence signal intensities by a factor > 5 compared to similar QDs grown without DBR mirrors, together with a remarkable reduction in the density of spectral lines and background. The optical characterization of the sample also revealed that the fraction of emitters with FSS < 50 $\mu$eV is approximately 50\%. Because of the wide applicability of the DE growth scheme, these results can be easily transferred to different wavelengths or material systems.
Finally, we observed that the QDs tend to have dipoles aligned along a specific direction, despite the C\textsubscript{3v} symmetry of the (111)-oriented substrate. In fact, numerical simulations based on the EPM and CI methods have confirmed that the presence of the miscut modifies the spatial distribution of the electron and hole wave functions, leading to their elongation in the [1$\bar{1}$0] direction. This phenomenon influences the polarization of the excitonic states introducing a clear preferential orientation in the natural C\textsubscript{3v} symmetry of the surface, while the addition of strain due to lattice mismatch modulates and enhances this effect.

In the future, linewidths and FSS could be reduced by changing the composition of the barrier layer from InAlAs to InGaAs \cite{Tuktamyshev.2021} and employing a smaller miscut angle, respectively. Moreover, the photon extraction efficiency could be further improved by replacing the DBR cavity with vertically emitting photonic structures, such as mesas \cite{Gao.2022}, micropillars \cite{Blokhin.2021, Bremer.2022} or circular Bragg gratings \cite{Rickert.2019, Barbiero.2022, Barbiero.2022b} in order to create efficient quantum light sources at the telecom wavelength for deployment and operation over the existing optical fiber network. 


\begin{acknowledgments}
\noindent This project has received partial funding from the European Union’s Horizon 2020 research and innovation programme under the Marie Skłodowska-Curie grant agreement No 721394. A. Barbiero thanks A. Tartakovskii for the academic supervision.

\end{acknowledgments}

\bibliography{paper_bibliography}

\begin{thebibliography}{48}%
\makeatletter
\providecommand \@ifxundefined [1]{%
 \@ifx{#1\undefined}
}%
\providecommand \@ifnum [1]{%
 \ifnum #1\expandafter \@firstoftwo
 \else \expandafter \@secondoftwo
 \fi
}%
\providecommand \@ifx [1]{%
 \ifx #1\expandafter \@firstoftwo
 \else \expandafter \@secondoftwo
 \fi
}%
\providecommand \natexlab [1]{#1}%
\providecommand \enquote  [1]{``#1''}%
\providecommand \bibnamefont  [1]{#1}%
\providecommand \bibfnamefont [1]{#1}%
\providecommand \citenamefont [1]{#1}%
\providecommand \href@noop [0]{\@secondoftwo}%
\providecommand \href [0]{\begingroup \@sanitize@url \@href}%
\providecommand \@href[1]{\@@startlink{#1}\@@href}%
\providecommand \@@href[1]{\endgroup#1\@@endlink}%
\providecommand \@sanitize@url [0]{\catcode `\\12\catcode `\$12\catcode
  `\&12\catcode `\#12\catcode `\^12\catcode `\_12\catcode `\%12\relax}%
\providecommand \@@startlink[1]{}%
\providecommand \@@endlink[0]{}%
\providecommand \url  [0]{\begingroup\@sanitize@url \@url }%
\providecommand \@url [1]{\endgroup\@href {#1}{\urlprefix }}%
\providecommand \urlprefix  [0]{URL }%
\providecommand \Eprint [0]{\href }%
\providecommand \doibase [0]{https://doi.org/}%
\providecommand \selectlanguage [0]{\@gobble}%
\providecommand \bibinfo  [0]{\@secondoftwo}%
\providecommand \bibfield  [0]{\@secondoftwo}%
\providecommand \translation [1]{[#1]}%
\providecommand \BibitemOpen [0]{}%
\providecommand \bibitemStop [0]{}%
\providecommand \bibitemNoStop [0]{.\EOS\space}%
\providecommand \EOS [0]{\spacefactor3000\relax}%
\providecommand \BibitemShut  [1]{\csname bibitem#1\endcsname}%
\let\auto@bib@innerbib\@empty
\bibitem [{\citenamefont {Kimble}(2008)}]{Kimble.2008}%
  \BibitemOpen
  \bibfield  {author} {\bibinfo {author} {\bibfnamefont {H.~J.}\ \bibnamefont
  {Kimble}},\ }\bibfield  {title} {\bibinfo {title} {The quantum internet},\
  }\href {https://doi.org/10.1038/nature07127} {\bibfield  {journal} {\bibinfo
  {journal} {Nature}\ }\textbf {\bibinfo {volume} {453}},\ \bibinfo {pages}
  {1023} (\bibinfo {year} {2008})}\BibitemShut {NoStop}%
\bibitem [{\citenamefont {Wehner}\ \emph {et~al.}(2018)\citenamefont {Wehner},
  \citenamefont {Elkouss},\ and\ \citenamefont {Hanson}}]{Wehner.2018}%
  \BibitemOpen
  \bibfield  {author} {\bibinfo {author} {\bibfnamefont {S.}~\bibnamefont
  {Wehner}}, \bibinfo {author} {\bibfnamefont {D.}~\bibnamefont {Elkouss}},\
  and\ \bibinfo {author} {\bibfnamefont {R.}~\bibnamefont {Hanson}},\
  }\bibfield  {title} {\bibinfo {title} {Quantum internet: A vision for the
  road ahead},\ }\bibfield  {journal} {\bibinfo  {journal} {Science (New York,
  N.Y.)}\ }\textbf {\bibinfo {volume} {362}},\ \href
  {https://doi.org/10.1126/science.aam9288} {10.1126/science.aam9288} (\bibinfo
  {year} {2018})\BibitemShut {NoStop}%
\bibitem [{\citenamefont {Skiba-Szymanska}\ \emph {et~al.}(2017)\citenamefont
  {Skiba-Szymanska}, \citenamefont {Stevenson}, \citenamefont {Varnava},
  \citenamefont {Felle}, \citenamefont {Huwer}, \citenamefont {M{\"u}ller},
  \citenamefont {Bennett}, \citenamefont {Lee}, \citenamefont {Farrer},
  \citenamefont {Krysa}, \citenamefont {Spencer}, \citenamefont {Goff},
  \citenamefont {Ritchie}, \citenamefont {Heffernan},\ and\ \citenamefont
  {Shields}}]{SkibaSzymanska.2017}%
  \BibitemOpen
  \bibfield  {author} {\bibinfo {author} {\bibfnamefont {J.}~\bibnamefont
  {Skiba-Szymanska}}, \bibinfo {author} {\bibfnamefont {R.~M.}\ \bibnamefont
  {Stevenson}}, \bibinfo {author} {\bibfnamefont {C.}~\bibnamefont {Varnava}},
  \bibinfo {author} {\bibfnamefont {M.}~\bibnamefont {Felle}}, \bibinfo
  {author} {\bibfnamefont {J.}~\bibnamefont {Huwer}}, \bibinfo {author}
  {\bibfnamefont {T.}~\bibnamefont {M{\"u}ller}}, \bibinfo {author}
  {\bibfnamefont {A.~J.}\ \bibnamefont {Bennett}}, \bibinfo {author}
  {\bibfnamefont {J.~P.}\ \bibnamefont {Lee}}, \bibinfo {author} {\bibfnamefont
  {I.}~\bibnamefont {Farrer}}, \bibinfo {author} {\bibfnamefont {A.~B.}\
  \bibnamefont {Krysa}}, \bibinfo {author} {\bibfnamefont {P.}~\bibnamefont
  {Spencer}}, \bibinfo {author} {\bibfnamefont {L.~E.}\ \bibnamefont {Goff}},
  \bibinfo {author} {\bibfnamefont {D.~A.}\ \bibnamefont {Ritchie}}, \bibinfo
  {author} {\bibfnamefont {J.}~\bibnamefont {Heffernan}},\ and\ \bibinfo
  {author} {\bibfnamefont {A.~J.}\ \bibnamefont {Shields}},\ }\bibfield
  {title} {\bibinfo {title} {Universal growth scheme for quantum dots with low
  fine-structure splitting at various emission wavelengths},\ }\bibfield
  {journal} {\bibinfo  {journal} {Physical Review Applied}\ }\textbf {\bibinfo
  {volume} {8}},\ \href {https://doi.org/10.1103/PhysRevApplied.8.014013}
  {10.1103/PhysRevApplied.8.014013} (\bibinfo {year} {2017})\BibitemShut
  {NoStop}%
\bibitem [{\citenamefont {Orieux}\ \emph {et~al.}(2017)\citenamefont {Orieux},
  \citenamefont {Versteegh}, \citenamefont {J{\"o}ns},\ and\ \citenamefont
  {Ducci}}]{Orieux.2017}%
  \BibitemOpen
  \bibfield  {author} {\bibinfo {author} {\bibfnamefont {A.}~\bibnamefont
  {Orieux}}, \bibinfo {author} {\bibfnamefont {M.~A.~M.}\ \bibnamefont
  {Versteegh}}, \bibinfo {author} {\bibfnamefont {K.~D.}\ \bibnamefont
  {J{\"o}ns}},\ and\ \bibinfo {author} {\bibfnamefont {S.}~\bibnamefont
  {Ducci}},\ }\bibfield  {title} {\bibinfo {title} {Semiconductor devices for
  entangled photon pair generation: a review},\ }\href
  {https://doi.org/10.1088/1361-6633/aa6955} {\bibfield  {journal} {\bibinfo
  {journal} {Reports on Progress in Physics}\ }\textbf {\bibinfo {volume}
  {80}},\ \bibinfo {pages} {076001} (\bibinfo {year} {2017})}\BibitemShut
  {NoStop}%
\bibitem [{\citenamefont {Huber}\ \emph {et~al.}(2018)\citenamefont {Huber},
  \citenamefont {Reindl}, \citenamefont {Aberl}, \citenamefont {Rastelli},\
  and\ \citenamefont {Trotta}}]{Huber.2018}%
  \BibitemOpen
  \bibfield  {author} {\bibinfo {author} {\bibfnamefont {D.}~\bibnamefont
  {Huber}}, \bibinfo {author} {\bibfnamefont {M.}~\bibnamefont {Reindl}},
  \bibinfo {author} {\bibfnamefont {J.}~\bibnamefont {Aberl}}, \bibinfo
  {author} {\bibfnamefont {A.}~\bibnamefont {Rastelli}},\ and\ \bibinfo
  {author} {\bibfnamefont {R.}~\bibnamefont {Trotta}},\ }\bibfield  {title}
  {\bibinfo {title} {Semiconductor quantum dots as an ideal source of
  polarization-entangled photon pairs on-demand: a review},\ }\href
  {https://doi.org/10.1088/2040-8986/aac4c4} {\bibfield  {journal} {\bibinfo
  {journal} {Japanese Journal of Applied Physics}\ }\textbf {\bibinfo {volume}
  {20}},\ \bibinfo {pages} {073002} (\bibinfo {year} {2018})}\BibitemShut
  {NoStop}%
\bibitem [{\citenamefont {Kir{\v{s}}ansk{\.{e}}}\ \emph
  {et~al.}(2017)\citenamefont {Kir{\v{s}}ansk{\.{e}}}, \citenamefont
  {Thyrrestrup}, \citenamefont {Daveau}, \citenamefont {Dree{\ss}en},
  \citenamefont {Pregnolato}, \citenamefont {Midolo}, \citenamefont
  {Tighineanu}, \citenamefont {Javadi}, \citenamefont {Stobbe}, \citenamefont
  {Schott}, \citenamefont {Ludwig}, \citenamefont {Wieck}, \citenamefont
  {Park}, \citenamefont {Song}, \citenamefont {Kuhlmann}, \citenamefont
  {S{\"o}llner}, \citenamefont {L{\"o}bl}, \citenamefont {Warburton},\ and\
  \citenamefont {Lodahl}}]{Kirsanske.2017}%
  \BibitemOpen
  \bibfield  {author} {\bibinfo {author} {\bibfnamefont {G.}~\bibnamefont
  {Kir{\v{s}}ansk{\.{e}}}}, \bibinfo {author} {\bibfnamefont {H.}~\bibnamefont
  {Thyrrestrup}}, \bibinfo {author} {\bibfnamefont {R.~S.}\ \bibnamefont
  {Daveau}}, \bibinfo {author} {\bibfnamefont {C.~L.}\ \bibnamefont
  {Dree{\ss}en}}, \bibinfo {author} {\bibfnamefont {T.}~\bibnamefont
  {Pregnolato}}, \bibinfo {author} {\bibfnamefont {L.}~\bibnamefont {Midolo}},
  \bibinfo {author} {\bibfnamefont {P.}~\bibnamefont {Tighineanu}}, \bibinfo
  {author} {\bibfnamefont {A.}~\bibnamefont {Javadi}}, \bibinfo {author}
  {\bibfnamefont {S.}~\bibnamefont {Stobbe}}, \bibinfo {author} {\bibfnamefont
  {R.}~\bibnamefont {Schott}}, \bibinfo {author} {\bibfnamefont
  {A.}~\bibnamefont {Ludwig}}, \bibinfo {author} {\bibfnamefont {A.~D.}\
  \bibnamefont {Wieck}}, \bibinfo {author} {\bibfnamefont {S.~I.}\ \bibnamefont
  {Park}}, \bibinfo {author} {\bibfnamefont {J.~D.}\ \bibnamefont {Song}},
  \bibinfo {author} {\bibfnamefont {A.~V.}\ \bibnamefont {Kuhlmann}}, \bibinfo
  {author} {\bibfnamefont {I.}~\bibnamefont {S{\"o}llner}}, \bibinfo {author}
  {\bibfnamefont {M.~C.}\ \bibnamefont {L{\"o}bl}}, \bibinfo {author}
  {\bibfnamefont {R.~J.}\ \bibnamefont {Warburton}},\ and\ \bibinfo {author}
  {\bibfnamefont {P.}~\bibnamefont {Lodahl}},\ }\bibfield  {title} {\bibinfo
  {title} {Indistinguishable and efficient single photons from a quantum dot in
  a planar nanobeam waveguide},\ }\href
  {https://doi.org/10.1103/PhysRevB.96.165306} {\bibfield  {journal} {\bibinfo
  {journal} {Physical Review B}\ }\textbf {\bibinfo {volume} {96}},\ \bibinfo
  {pages} {174} (\bibinfo {year} {2017})}\BibitemShut {NoStop}%
\bibitem [{\citenamefont {Uppu}\ \emph {et~al.}(2020)\citenamefont {Uppu},
  \citenamefont {Pedersen}, \citenamefont {Wang}, \citenamefont {Olesen},
  \citenamefont {Papon}, \citenamefont {Zhou}, \citenamefont {Midolo},
  \citenamefont {Scholz}, \citenamefont {Wieck}, \citenamefont {Ludwig},\ and\
  \citenamefont {Lodahl}}]{Uppu.2020}%
  \BibitemOpen
  \bibfield  {author} {\bibinfo {author} {\bibfnamefont {R.}~\bibnamefont
  {Uppu}}, \bibinfo {author} {\bibfnamefont {F.~T.}\ \bibnamefont {Pedersen}},
  \bibinfo {author} {\bibfnamefont {Y.}~\bibnamefont {Wang}}, \bibinfo {author}
  {\bibfnamefont {C.~T.}\ \bibnamefont {Olesen}}, \bibinfo {author}
  {\bibfnamefont {C.}~\bibnamefont {Papon}}, \bibinfo {author} {\bibfnamefont
  {X.}~\bibnamefont {Zhou}}, \bibinfo {author} {\bibfnamefont {L.}~\bibnamefont
  {Midolo}}, \bibinfo {author} {\bibfnamefont {S.}~\bibnamefont {Scholz}},
  \bibinfo {author} {\bibfnamefont {A.~D.}\ \bibnamefont {Wieck}}, \bibinfo
  {author} {\bibfnamefont {A.}~\bibnamefont {Ludwig}},\ and\ \bibinfo {author}
  {\bibfnamefont {P.}~\bibnamefont {Lodahl}},\ }\bibfield  {title} {\bibinfo
  {title} {Scalable integrated single-photon source},\ }\bibfield  {journal}
  {\bibinfo  {journal} {Science advances}\ }\textbf {\bibinfo {volume} {6}},\
  \href {https://doi.org/10.1126/sciadv.abc8268} {10.1126/sciadv.abc8268}
  (\bibinfo {year} {2020})\BibitemShut {NoStop}%
\bibitem [{\citenamefont {Thomas}\ \emph {et~al.}(2021)\citenamefont {Thomas},
  \citenamefont {Billard}, \citenamefont {Coste}, \citenamefont {Wein},
  \citenamefont {Priya}, \citenamefont {Ollivier}, \citenamefont {Krebs},
  \citenamefont {Taza{\"i}rt}, \citenamefont {Harouri}, \citenamefont
  {Lemaitre}, \citenamefont {Sagnes}, \citenamefont {Anton}, \citenamefont
  {Lanco}, \citenamefont {Somaschi}, \citenamefont {Loredo},\ and\
  \citenamefont {Senellart}}]{Thomas.2021b}%
  \BibitemOpen
  \bibfield  {author} {\bibinfo {author} {\bibfnamefont {S.~E.}\ \bibnamefont
  {Thomas}}, \bibinfo {author} {\bibfnamefont {M.}~\bibnamefont {Billard}},
  \bibinfo {author} {\bibfnamefont {N.}~\bibnamefont {Coste}}, \bibinfo
  {author} {\bibfnamefont {S.~C.}\ \bibnamefont {Wein}}, \bibinfo {author}
  {\bibnamefont {Priya}}, \bibinfo {author} {\bibfnamefont {H.}~\bibnamefont
  {Ollivier}}, \bibinfo {author} {\bibfnamefont {O.}~\bibnamefont {Krebs}},
  \bibinfo {author} {\bibfnamefont {L.}~\bibnamefont {Taza{\"i}rt}}, \bibinfo
  {author} {\bibfnamefont {A.}~\bibnamefont {Harouri}}, \bibinfo {author}
  {\bibfnamefont {A.}~\bibnamefont {Lemaitre}}, \bibinfo {author}
  {\bibfnamefont {I.}~\bibnamefont {Sagnes}}, \bibinfo {author} {\bibfnamefont
  {C.}~\bibnamefont {Anton}}, \bibinfo {author} {\bibfnamefont
  {L.}~\bibnamefont {Lanco}}, \bibinfo {author} {\bibfnamefont
  {N.}~\bibnamefont {Somaschi}}, \bibinfo {author} {\bibfnamefont {J.~C.}\
  \bibnamefont {Loredo}},\ and\ \bibinfo {author} {\bibfnamefont
  {P.}~\bibnamefont {Senellart}},\ }\bibfield  {title} {\bibinfo {title}
  {Bright polarized single-photon source based on a linear dipole},\ }\href
  {https://doi.org/10.1103/PhysRevLett.126.233601} {\bibfield  {journal}
  {\bibinfo  {journal} {Physical review letters}\ }\textbf {\bibinfo {volume}
  {126}},\ \bibinfo {pages} {233601} (\bibinfo {year} {2021})}\BibitemShut
  {NoStop}%
\bibitem [{\citenamefont {Tomm}\ \emph {et~al.}(2021)\citenamefont {Tomm},
  \citenamefont {Javadi}, \citenamefont {Antoniadis}, \citenamefont {Najer},
  \citenamefont {L{\"o}bl}, \citenamefont {Korsch}, \citenamefont {Schott},
  \citenamefont {Valentin}, \citenamefont {Wieck}, \citenamefont {Ludwig},\
  and\ \citenamefont {Warburton}}]{Tomm.2021}%
  \BibitemOpen
  \bibfield  {author} {\bibinfo {author} {\bibfnamefont {N.}~\bibnamefont
  {Tomm}}, \bibinfo {author} {\bibfnamefont {A.}~\bibnamefont {Javadi}},
  \bibinfo {author} {\bibfnamefont {N.~O.}\ \bibnamefont {Antoniadis}},
  \bibinfo {author} {\bibfnamefont {D.}~\bibnamefont {Najer}}, \bibinfo
  {author} {\bibfnamefont {M.~C.}\ \bibnamefont {L{\"o}bl}}, \bibinfo {author}
  {\bibfnamefont {A.~R.}\ \bibnamefont {Korsch}}, \bibinfo {author}
  {\bibfnamefont {R.}~\bibnamefont {Schott}}, \bibinfo {author} {\bibfnamefont
  {S.~R.}\ \bibnamefont {Valentin}}, \bibinfo {author} {\bibfnamefont {A.~D.}\
  \bibnamefont {Wieck}}, \bibinfo {author} {\bibfnamefont {A.}~\bibnamefont
  {Ludwig}},\ and\ \bibinfo {author} {\bibfnamefont {R.~J.}\ \bibnamefont
  {Warburton}},\ }\bibfield  {title} {\bibinfo {title} {A bright and fast
  source of coherent single photons},\ }\bibfield  {journal} {\bibinfo
  {journal} {Nature Nanotechnology}\ }\href
  {https://doi.org/10.1038/s41565-020-00831-x} {10.1038/s41565-020-00831-x}
  (\bibinfo {year} {2021})\BibitemShut {NoStop}%
\bibitem [{\citenamefont {Sartison}\ \emph {et~al.}(2017)\citenamefont
  {Sartison}, \citenamefont {Portalupi}, \citenamefont {Gissibl}, \citenamefont
  {Jetter}, \citenamefont {Giessen},\ and\ \citenamefont
  {Michler}}]{Sartison.2017}%
  \BibitemOpen
  \bibfield  {author} {\bibinfo {author} {\bibfnamefont {M.}~\bibnamefont
  {Sartison}}, \bibinfo {author} {\bibfnamefont {S.~L.}\ \bibnamefont
  {Portalupi}}, \bibinfo {author} {\bibfnamefont {T.}~\bibnamefont {Gissibl}},
  \bibinfo {author} {\bibfnamefont {M.}~\bibnamefont {Jetter}}, \bibinfo
  {author} {\bibfnamefont {H.}~\bibnamefont {Giessen}},\ and\ \bibinfo {author}
  {\bibfnamefont {P.}~\bibnamefont {Michler}},\ }\bibfield  {title} {\bibinfo
  {title} {Combining in-situ lithography with 3d printed solid immersion lenses
  for single quantum dot spectroscopy},\ }\href
  {https://doi.org/10.1038/srep39916} {\bibfield  {journal} {\bibinfo
  {journal} {Scientific Reports}\ }\textbf {\bibinfo {volume} {7}},\ \bibinfo
  {pages} {39916} (\bibinfo {year} {2017})}\BibitemShut {NoStop}%
\bibitem [{\citenamefont {Chen}\ \emph {et~al.}(2018)\citenamefont {Chen},
  \citenamefont {Zopf}, \citenamefont {Keil}, \citenamefont {Ding},\ and\
  \citenamefont {Schmidt}}]{Chen.2018}%
  \BibitemOpen
  \bibfield  {author} {\bibinfo {author} {\bibfnamefont {Y.}~\bibnamefont
  {Chen}}, \bibinfo {author} {\bibfnamefont {M.}~\bibnamefont {Zopf}}, \bibinfo
  {author} {\bibfnamefont {R.}~\bibnamefont {Keil}}, \bibinfo {author}
  {\bibfnamefont {F.}~\bibnamefont {Ding}},\ and\ \bibinfo {author}
  {\bibfnamefont {O.~G.}\ \bibnamefont {Schmidt}},\ }\bibfield  {title}
  {\bibinfo {title} {Highly-efficient extraction of entangled photons from
  quantum dots using a broadband optical antenna},\ }\href
  {https://doi.org/10.1038/s41467-018-05456-2} {\bibfield  {journal} {\bibinfo
  {journal} {Nature communications}\ }\textbf {\bibinfo {volume} {9}},\
  \bibinfo {pages} {2994} (\bibinfo {year} {2018})}\BibitemShut {NoStop}%
\bibitem [{\citenamefont {Kaganskiy}\ \emph {et~al.}(2018)\citenamefont
  {Kaganskiy}, \citenamefont {Fischbach}, \citenamefont {Strittmatter},
  \citenamefont {Heindel}, \citenamefont {Rodt},\ and\ \citenamefont
  {Reitzenstein}}]{Kaganskiy.2018}%
  \BibitemOpen
  \bibfield  {author} {\bibinfo {author} {\bibfnamefont {A.}~\bibnamefont
  {Kaganskiy}}, \bibinfo {author} {\bibfnamefont {S.}~\bibnamefont
  {Fischbach}}, \bibinfo {author} {\bibfnamefont {A.}~\bibnamefont
  {Strittmatter}}, \bibinfo {author} {\bibfnamefont {T.}~\bibnamefont
  {Heindel}}, \bibinfo {author} {\bibfnamefont {S.}~\bibnamefont {Rodt}},\ and\
  \bibinfo {author} {\bibfnamefont {S.}~\bibnamefont {Reitzenstein}},\
  }\bibfield  {title} {\bibinfo {title} {Enhancing the photon-extraction
  efficiency of site-controlled quantum dots by deterministically fabricated
  microlenses},\ }\href {https://doi.org/10.1016/j.optcom.2017.12.032}
  {\bibfield  {journal} {\bibinfo  {journal} {Optics Communications}\ }\textbf
  {\bibinfo {volume} {413}},\ \bibinfo {pages} {162} (\bibinfo {year}
  {2018})}\BibitemShut {NoStop}%
\bibitem [{\citenamefont {Gammon}\ \emph {et~al.}(1996)\citenamefont {Gammon},
  \citenamefont {Snow}, \citenamefont {Shanabrook}, \citenamefont {Katzer},\
  and\ \citenamefont {Park}}]{Gammon.1996}%
  \BibitemOpen
  \bibfield  {author} {\bibinfo {author} {\bibnamefont {Gammon}}, \bibinfo
  {author} {\bibnamefont {Snow}}, \bibinfo {author} {\bibnamefont
  {Shanabrook}}, \bibinfo {author} {\bibnamefont {Katzer}},\ and\ \bibinfo
  {author} {\bibnamefont {Park}},\ }\bibfield  {title} {\bibinfo {title} {Fine
  structure splitting in the optical spectra of single {G}a{A}s quantum dots},\
  }\href {https://doi.org/10.1103/PhysRevLett.76.3005} {\bibfield  {journal}
  {\bibinfo  {journal} {Physical review letters}\ }\textbf {\bibinfo {volume}
  {76}},\ \bibinfo {pages} {3005} (\bibinfo {year} {1996})}\BibitemShut
  {NoStop}%
\bibitem [{\citenamefont {Kulakovskii}\ \emph {et~al.}(1999)\citenamefont
  {Kulakovskii}, \citenamefont {Bacher}, \citenamefont {Weigand}, \citenamefont
  {K{\"u}mmell}, \citenamefont {Forchel}, \citenamefont {Borovitskaya},
  \citenamefont {Leonardi},\ and\ \citenamefont {Hommel}}]{Kulakovskii.1999}%
  \BibitemOpen
  \bibfield  {author} {\bibinfo {author} {\bibfnamefont {V.~D.}\ \bibnamefont
  {Kulakovskii}}, \bibinfo {author} {\bibfnamefont {G.}~\bibnamefont {Bacher}},
  \bibinfo {author} {\bibfnamefont {R.}~\bibnamefont {Weigand}}, \bibinfo
  {author} {\bibfnamefont {T.}~\bibnamefont {K{\"u}mmell}}, \bibinfo {author}
  {\bibfnamefont {A.}~\bibnamefont {Forchel}}, \bibinfo {author} {\bibfnamefont
  {E.}~\bibnamefont {Borovitskaya}}, \bibinfo {author} {\bibfnamefont
  {K.}~\bibnamefont {Leonardi}},\ and\ \bibinfo {author} {\bibfnamefont
  {D.}~\bibnamefont {Hommel}},\ }\bibfield  {title} {\bibinfo {title} {Fine
  structure of biexciton emission in symmetric and asymmetric {C}d{S}e/{Z}n{S}e
  single quantum dots},\ }\href {https://doi.org/10.1103/PhysRevLett.82.1780}
  {\bibfield  {journal} {\bibinfo  {journal} {Physical review letters}\
  }\textbf {\bibinfo {volume} {82}},\ \bibinfo {pages} {1780} (\bibinfo {year}
  {1999})}\BibitemShut {NoStop}%
\bibitem [{\citenamefont {Mano}\ \emph {et~al.}(2010)\citenamefont {Mano},
  \citenamefont {Abbarchi}, \citenamefont {Kuroda}, \citenamefont {McSkimming},
  \citenamefont {Ohtake}, \citenamefont {Mitsuishi},\ and\ \citenamefont
  {Sakoda}}]{Mano.2010}%
  \BibitemOpen
  \bibfield  {author} {\bibinfo {author} {\bibfnamefont {T.}~\bibnamefont
  {Mano}}, \bibinfo {author} {\bibfnamefont {M.}~\bibnamefont {Abbarchi}},
  \bibinfo {author} {\bibfnamefont {T.}~\bibnamefont {Kuroda}}, \bibinfo
  {author} {\bibfnamefont {B.}~\bibnamefont {McSkimming}}, \bibinfo {author}
  {\bibfnamefont {A.}~\bibnamefont {Ohtake}}, \bibinfo {author} {\bibfnamefont
  {K.}~\bibnamefont {Mitsuishi}},\ and\ \bibinfo {author} {\bibfnamefont
  {K.}~\bibnamefont {Sakoda}},\ }\bibfield  {title} {\bibinfo {title}
  {Self-assembly of symmetric {G}a{A}s quantum dots on (111){A} substrates:
  Suppression of fine-structure splitting},\ }\href
  {https://doi.org/10.1143/APEX.3.065203} {\bibfield  {journal} {\bibinfo
  {journal} {Applied Physics Express}\ }\textbf {\bibinfo {volume} {3}},\
  \bibinfo {pages} {065203} (\bibinfo {year} {2010})}\BibitemShut {NoStop}%
\bibitem [{\citenamefont {Kuroda}\ \emph {et~al.}(2013)\citenamefont {Kuroda},
  \citenamefont {Mano}, \citenamefont {Ha}, \citenamefont {Nakajima},
  \citenamefont {Kumano}, \citenamefont {Urbaszek}, \citenamefont {Jo},
  \citenamefont {Abbarchi}, \citenamefont {Sakuma}, \citenamefont {Sakoda},
  \citenamefont {Suemune}, \citenamefont {Marie},\ and\ \citenamefont
  {Amand}}]{Kuroda.2013}%
  \BibitemOpen
  \bibfield  {author} {\bibinfo {author} {\bibfnamefont {T.}~\bibnamefont
  {Kuroda}}, \bibinfo {author} {\bibfnamefont {T.}~\bibnamefont {Mano}},
  \bibinfo {author} {\bibfnamefont {N.}~\bibnamefont {Ha}}, \bibinfo {author}
  {\bibfnamefont {H.}~\bibnamefont {Nakajima}}, \bibinfo {author}
  {\bibfnamefont {H.}~\bibnamefont {Kumano}}, \bibinfo {author} {\bibfnamefont
  {B.}~\bibnamefont {Urbaszek}}, \bibinfo {author} {\bibfnamefont
  {M.}~\bibnamefont {Jo}}, \bibinfo {author} {\bibfnamefont {M.}~\bibnamefont
  {Abbarchi}}, \bibinfo {author} {\bibfnamefont {Y.}~\bibnamefont {Sakuma}},
  \bibinfo {author} {\bibfnamefont {K.}~\bibnamefont {Sakoda}}, \bibinfo
  {author} {\bibfnamefont {I.}~\bibnamefont {Suemune}}, \bibinfo {author}
  {\bibfnamefont {X.}~\bibnamefont {Marie}},\ and\ \bibinfo {author}
  {\bibfnamefont {T.}~\bibnamefont {Amand}},\ }\bibfield  {title} {\bibinfo
  {title} {Symmetric quantum dots as efficient sources of highly entangled
  photons: Violation of bell's inequality without spectral and temporal
  filtering},\ }\bibfield  {journal} {\bibinfo  {journal} {Physical Review B}\
  }\textbf {\bibinfo {volume} {88}},\ \href
  {https://doi.org/10.1103/PhysRevB.88.041306} {10.1103/PhysRevB.88.041306}
  (\bibinfo {year} {2013})\BibitemShut {NoStop}%
\bibitem [{\citenamefont {{Basso Basset}}\ \emph {et~al.}(2018)\citenamefont
  {{Basso Basset}}, \citenamefont {Bietti}, \citenamefont {Reindl},
  \citenamefont {Esposito}, \citenamefont {Fedorov}, \citenamefont {Huber},
  \citenamefont {Rastelli}, \citenamefont {Bonera}, \citenamefont {Trotta},\
  and\ \citenamefont {Sanguinetti}}]{BassoBasset.2018}%
  \BibitemOpen
  \bibfield  {author} {\bibinfo {author} {\bibfnamefont {F.}~\bibnamefont
  {{Basso Basset}}}, \bibinfo {author} {\bibfnamefont {S.}~\bibnamefont
  {Bietti}}, \bibinfo {author} {\bibfnamefont {M.}~\bibnamefont {Reindl}},
  \bibinfo {author} {\bibfnamefont {L.}~\bibnamefont {Esposito}}, \bibinfo
  {author} {\bibfnamefont {A.}~\bibnamefont {Fedorov}}, \bibinfo {author}
  {\bibfnamefont {D.}~\bibnamefont {Huber}}, \bibinfo {author} {\bibfnamefont
  {A.}~\bibnamefont {Rastelli}}, \bibinfo {author} {\bibfnamefont
  {E.}~\bibnamefont {Bonera}}, \bibinfo {author} {\bibfnamefont
  {R.}~\bibnamefont {Trotta}},\ and\ \bibinfo {author} {\bibfnamefont
  {S.}~\bibnamefont {Sanguinetti}},\ }\bibfield  {title} {\bibinfo {title}
  {High-yield fabrication of entangled photon emitters for hybrid quantum
  networking using high-temperature droplet epitaxy},\ }\href
  {https://doi.org/10.1021/acs.nanolett.7b04472} {\bibfield  {journal}
  {\bibinfo  {journal} {Nano letters}\ }\textbf {\bibinfo {volume} {18}},\
  \bibinfo {pages} {505} (\bibinfo {year} {2018})}\BibitemShut {NoStop}%
\bibitem [{\citenamefont {Ha}\ \emph {et~al.}(2015)\citenamefont {Ha},
  \citenamefont {Mano}, \citenamefont {Kuroda}, \citenamefont {Mitsuishi},
  \citenamefont {Ohtake}, \citenamefont {Castellano}, \citenamefont
  {Sanguinetti}, \citenamefont {Noda}, \citenamefont {Sakuma},\ and\
  \citenamefont {Sakoda}}]{Ha.2015}%
  \BibitemOpen
  \bibfield  {author} {\bibinfo {author} {\bibfnamefont {N.}~\bibnamefont
  {Ha}}, \bibinfo {author} {\bibfnamefont {T.}~\bibnamefont {Mano}}, \bibinfo
  {author} {\bibfnamefont {T.}~\bibnamefont {Kuroda}}, \bibinfo {author}
  {\bibfnamefont {K.}~\bibnamefont {Mitsuishi}}, \bibinfo {author}
  {\bibfnamefont {A.}~\bibnamefont {Ohtake}}, \bibinfo {author} {\bibfnamefont
  {A.}~\bibnamefont {Castellano}}, \bibinfo {author} {\bibfnamefont
  {S.}~\bibnamefont {Sanguinetti}}, \bibinfo {author} {\bibfnamefont
  {T.}~\bibnamefont {Noda}}, \bibinfo {author} {\bibfnamefont {Y.}~\bibnamefont
  {Sakuma}},\ and\ \bibinfo {author} {\bibfnamefont {K.}~\bibnamefont
  {Sakoda}},\ }\bibfield  {title} {\bibinfo {title} {Droplet epitaxy growth of
  telecom {I}n{A}s quantum dots on metamorphic {I}n{A}l{A}s/{G}a{A}s(111){A}},\
  }\href {https://doi.org/10.7567/JJAP.54.04DH07} {\bibfield  {journal}
  {\bibinfo  {journal} {Japanese Journal of Applied Physics}\ }\textbf
  {\bibinfo {volume} {54}},\ \bibinfo {pages} {04DH07} (\bibinfo {year}
  {2015})}\BibitemShut {NoStop}%
\bibitem [{\citenamefont {Yamaguchi}\ \emph {et~al.}(1997)\citenamefont
  {Yamaguchi}, \citenamefont {Belk}, \citenamefont {Zhang}, \citenamefont
  {Sudijono}, \citenamefont {Fahy}, \citenamefont {Jones}, \citenamefont
  {Pashley},\ and\ \citenamefont {Joyce}}]{Yamaguchi.1997}%
  \BibitemOpen
  \bibfield  {author} {\bibinfo {author} {\bibfnamefont {H.}~\bibnamefont
  {Yamaguchi}}, \bibinfo {author} {\bibfnamefont {J.~G.}\ \bibnamefont {Belk}},
  \bibinfo {author} {\bibfnamefont {X.~M.}\ \bibnamefont {Zhang}}, \bibinfo
  {author} {\bibfnamefont {J.~L.}\ \bibnamefont {Sudijono}}, \bibinfo {author}
  {\bibfnamefont {M.~R.}\ \bibnamefont {Fahy}}, \bibinfo {author}
  {\bibfnamefont {T.~S.}\ \bibnamefont {Jones}}, \bibinfo {author}
  {\bibfnamefont {D.~W.}\ \bibnamefont {Pashley}},\ and\ \bibinfo {author}
  {\bibfnamefont {B.~A.}\ \bibnamefont {Joyce}},\ }\bibfield  {title} {\bibinfo
  {title} {Atomic-scale imaging of strain relaxation via misfit dislocations in
  highly mismatched semiconductor heteroepitaxy: {I}n{A}s/{G}a{A}s(111){A}},\
  }\href {https://doi.org/10.1103/PhysRevB.55.1337} {\bibfield  {journal}
  {\bibinfo  {journal} {Physical review. B, Condensed matter}\ }\textbf
  {\bibinfo {volume} {55}},\ \bibinfo {pages} {1337} (\bibinfo {year}
  {1997})}\BibitemShut {NoStop}%
\bibitem [{\citenamefont {Koguchi}\ \emph {et~al.}(1991)\citenamefont
  {Koguchi}, \citenamefont {Takahashi},\ and\ \citenamefont
  {Chikyow}}]{Koguchi.1991}%
  \BibitemOpen
  \bibfield  {author} {\bibinfo {author} {\bibfnamefont {N.}~\bibnamefont
  {Koguchi}}, \bibinfo {author} {\bibfnamefont {S.}~\bibnamefont {Takahashi}},\
  and\ \bibinfo {author} {\bibfnamefont {T.}~\bibnamefont {Chikyow}},\
  }\bibfield  {title} {\bibinfo {title} {New {MBE} growth method for {I}n{S}b
  quantum well boxes},\ }\href {https://doi.org/10.1016/0022-0248(91)91064-H}
  {\bibfield  {journal} {\bibinfo  {journal} {Journal of Crystal Growth}\
  }\textbf {\bibinfo {volume} {111}},\ \bibinfo {pages} {688–692} (\bibinfo
  {year} {1991})}\BibitemShut {NoStop}%
\bibitem [{\citenamefont {Gurioli}\ \emph {et~al.}(2019)\citenamefont
  {Gurioli}, \citenamefont {Wang}, \citenamefont {Rastelli}, \citenamefont
  {Kuroda},\ and\ \citenamefont {Sanguinetti}}]{Gurioli.2019}%
  \BibitemOpen
  \bibfield  {author} {\bibinfo {author} {\bibfnamefont {M.}~\bibnamefont
  {Gurioli}}, \bibinfo {author} {\bibfnamefont {Z.}~\bibnamefont {Wang}},
  \bibinfo {author} {\bibfnamefont {A.}~\bibnamefont {Rastelli}}, \bibinfo
  {author} {\bibfnamefont {T.}~\bibnamefont {Kuroda}},\ and\ \bibinfo {author}
  {\bibfnamefont {S.}~\bibnamefont {Sanguinetti}},\ }\bibfield  {title}
  {\bibinfo {title} {Droplet epitaxy of semiconductor nanostructures for
  quantum photonic devices},\ }\href
  {https://doi.org/10.1038/s41563-019-0355-y} {\bibfield  {journal} {\bibinfo
  {journal} {Nature Materials}\ }\textbf {\bibinfo {volume} {18}},\ \bibinfo
  {pages} {799} (\bibinfo {year} {2019})}\BibitemShut {NoStop}%
\bibitem [{\citenamefont {Somaschini}\ \emph {et~al.}(2009)\citenamefont
  {Somaschini}, \citenamefont {Bietti}, \citenamefont {Koguchi},\ and\
  \citenamefont {Sanguinetti}}]{Somaschini.2009}%
  \BibitemOpen
  \bibfield  {author} {\bibinfo {author} {\bibfnamefont {C.}~\bibnamefont
  {Somaschini}}, \bibinfo {author} {\bibfnamefont {S.}~\bibnamefont {Bietti}},
  \bibinfo {author} {\bibfnamefont {N.}~\bibnamefont {Koguchi}},\ and\ \bibinfo
  {author} {\bibfnamefont {S.}~\bibnamefont {Sanguinetti}},\ }\bibfield
  {title} {\bibinfo {title} {Fabrication of multiple concentric nanoring
  structures},\ }\href {https://doi.org/10.1021/nl901493f} {\bibfield
  {journal} {\bibinfo  {journal} {Nano letters}\ }\textbf {\bibinfo {volume}
  {9}},\ \bibinfo {pages} {3419} (\bibinfo {year} {2009})}\BibitemShut
  {NoStop}%
\bibitem [{\citenamefont {Somaschini}\ \emph {et~al.}(2010)\citenamefont
  {Somaschini}, \citenamefont {Bietti}, \citenamefont {Koguchi},\ and\
  \citenamefont {Sanguinetti}}]{Somaschini.2010}%
  \BibitemOpen
  \bibfield  {author} {\bibinfo {author} {\bibfnamefont {C.}~\bibnamefont
  {Somaschini}}, \bibinfo {author} {\bibfnamefont {S.}~\bibnamefont {Bietti}},
  \bibinfo {author} {\bibfnamefont {N.}~\bibnamefont {Koguchi}},\ and\ \bibinfo
  {author} {\bibfnamefont {S.}~\bibnamefont {Sanguinetti}},\ }\bibfield
  {title} {\bibinfo {title} {Shape control via surface reconstruction kinetics
  of droplet epitaxy nanostructures},\ }\href
  {https://doi.org/10.1063/1.3511283} {\bibfield  {journal} {\bibinfo
  {journal} {Applied Physics Letters}\ }\textbf {\bibinfo {volume} {97}},\
  \bibinfo {pages} {203109} (\bibinfo {year} {2010})}\BibitemShut {NoStop}%
\bibitem [{\citenamefont {Tuktamyshev}\ \emph {et~al.}(2021)\citenamefont
  {Tuktamyshev}, \citenamefont {Fedorov}, \citenamefont {Bietti}, \citenamefont
  {Vichi}, \citenamefont {Zeuner}, \citenamefont {J{\"o}ns}, \citenamefont
  {Chrastina}, \citenamefont {Tsukamoto}, \citenamefont {Zwiller},
  \citenamefont {Gurioli},\ and\ \citenamefont
  {Sanguinetti}}]{Tuktamyshev.2021}%
  \BibitemOpen
  \bibfield  {author} {\bibinfo {author} {\bibfnamefont {A.}~\bibnamefont
  {Tuktamyshev}}, \bibinfo {author} {\bibfnamefont {A.}~\bibnamefont
  {Fedorov}}, \bibinfo {author} {\bibfnamefont {S.}~\bibnamefont {Bietti}},
  \bibinfo {author} {\bibfnamefont {S.}~\bibnamefont {Vichi}}, \bibinfo
  {author} {\bibfnamefont {K.~D.}\ \bibnamefont {Zeuner}}, \bibinfo {author}
  {\bibfnamefont {K.~D.}\ \bibnamefont {J{\"o}ns}}, \bibinfo {author}
  {\bibfnamefont {D.}~\bibnamefont {Chrastina}}, \bibinfo {author}
  {\bibfnamefont {S.}~\bibnamefont {Tsukamoto}}, \bibinfo {author}
  {\bibfnamefont {V.}~\bibnamefont {Zwiller}}, \bibinfo {author} {\bibfnamefont
  {M.}~\bibnamefont {Gurioli}},\ and\ \bibinfo {author} {\bibfnamefont
  {S.}~\bibnamefont {Sanguinetti}},\ }\bibfield  {title} {\bibinfo {title}
  {Telecom-wavelength inas qds with low fine structure splitting grown by
  droplet epitaxy on {G}a{A}s(111){A} vicinal substrates},\ }\href
  {https://doi.org/10.1063/5.0045776} {\bibfield  {journal} {\bibinfo
  {journal} {Applied Physics Letters}\ }\textbf {\bibinfo {volume} {118}},\
  \bibinfo {pages} {133102} (\bibinfo {year} {2021})}\BibitemShut {NoStop}%
\bibitem [{\citenamefont {Ward}\ \emph {et~al.}(2014)\citenamefont {Ward},
  \citenamefont {Dean}, \citenamefont {Stevenson}, \citenamefont {Bennett},
  \citenamefont {Ellis}, \citenamefont {Cooper}, \citenamefont {Farrer},
  \citenamefont {Nicoll}, \citenamefont {Ritchie},\ and\ \citenamefont
  {Shields}}]{Ward.2014}%
  \BibitemOpen
  \bibfield  {author} {\bibinfo {author} {\bibfnamefont {M.~B.}\ \bibnamefont
  {Ward}}, \bibinfo {author} {\bibfnamefont {M.~C.}\ \bibnamefont {Dean}},
  \bibinfo {author} {\bibfnamefont {R.~M.}\ \bibnamefont {Stevenson}}, \bibinfo
  {author} {\bibfnamefont {A.~J.}\ \bibnamefont {Bennett}}, \bibinfo {author}
  {\bibfnamefont {D.}~\bibnamefont {Ellis}}, \bibinfo {author} {\bibfnamefont
  {K.}~\bibnamefont {Cooper}}, \bibinfo {author} {\bibfnamefont
  {I.}~\bibnamefont {Farrer}}, \bibinfo {author} {\bibfnamefont {C.~A.}\
  \bibnamefont {Nicoll}}, \bibinfo {author} {\bibfnamefont {D.~A.}\
  \bibnamefont {Ritchie}},\ and\ \bibinfo {author} {\bibfnamefont {A.~J.}\
  \bibnamefont {Shields}},\ }\bibfield  {title} {\bibinfo {title} {Coherent
  dynamics of a telecom-wavelength entangled photon source},\ }\href
  {https://doi.org/10.1038/ncomms4316} {\bibfield  {journal} {\bibinfo
  {journal} {Nature communications}\ }\textbf {\bibinfo {volume} {5}},\
  \bibinfo {pages} {3316} (\bibinfo {year} {2014})}\BibitemShut {NoStop}%
\bibitem [{\citenamefont {Huwer}\ \emph {et~al.}(2017)\citenamefont {Huwer},
  \citenamefont {Stevenson}, \citenamefont {Skiba-Szymanska}, \citenamefont
  {Ward}, \citenamefont {Shields}, \citenamefont {Felle}, \citenamefont
  {Farrer}, \citenamefont {Ritchie},\ and\ \citenamefont {Penty}}]{Huwer.2017}%
  \BibitemOpen
  \bibfield  {author} {\bibinfo {author} {\bibfnamefont {J.}~\bibnamefont
  {Huwer}}, \bibinfo {author} {\bibfnamefont {R.~M.}\ \bibnamefont
  {Stevenson}}, \bibinfo {author} {\bibfnamefont {J.}~\bibnamefont
  {Skiba-Szymanska}}, \bibinfo {author} {\bibfnamefont {M.~B.}\ \bibnamefont
  {Ward}}, \bibinfo {author} {\bibfnamefont {A.~J.}\ \bibnamefont {Shields}},
  \bibinfo {author} {\bibfnamefont {M.}~\bibnamefont {Felle}}, \bibinfo
  {author} {\bibfnamefont {I.}~\bibnamefont {Farrer}}, \bibinfo {author}
  {\bibfnamefont {D.~A.}\ \bibnamefont {Ritchie}},\ and\ \bibinfo {author}
  {\bibfnamefont {R.~V.}\ \bibnamefont {Penty}},\ }\bibfield  {title} {\bibinfo
  {title} {Quantum-dot-based telecommunication-wavelength quantum relay},\
  }\bibfield  {journal} {\bibinfo  {journal} {Physical Review Applied}\
  }\textbf {\bibinfo {volume} {8}},\ \href
  {https://doi.org/10.1103/PhysRevApplied.8.024007}
  {10.1103/PhysRevApplied.8.024007} (\bibinfo {year} {2017})\BibitemShut
  {NoStop}%
\bibitem [{\citenamefont {M{\"u}ller}\ \emph {et~al.}(2018)\citenamefont
  {M{\"u}ller}, \citenamefont {Skiba-Szymanska}, \citenamefont {Krysa},
  \citenamefont {Huwer}, \citenamefont {Felle}, \citenamefont {Anderson},
  \citenamefont {Stevenson}, \citenamefont {Heffernan}, \citenamefont
  {Ritchie},\ and\ \citenamefont {Shields}}]{Muller.2018}%
  \BibitemOpen
  \bibfield  {author} {\bibinfo {author} {\bibfnamefont {T.}~\bibnamefont
  {M{\"u}ller}}, \bibinfo {author} {\bibfnamefont {J.}~\bibnamefont
  {Skiba-Szymanska}}, \bibinfo {author} {\bibfnamefont {A.~B.}\ \bibnamefont
  {Krysa}}, \bibinfo {author} {\bibfnamefont {J.}~\bibnamefont {Huwer}},
  \bibinfo {author} {\bibfnamefont {M.}~\bibnamefont {Felle}}, \bibinfo
  {author} {\bibfnamefont {M.}~\bibnamefont {Anderson}}, \bibinfo {author}
  {\bibfnamefont {R.~M.}\ \bibnamefont {Stevenson}}, \bibinfo {author}
  {\bibfnamefont {J.}~\bibnamefont {Heffernan}}, \bibinfo {author}
  {\bibfnamefont {D.~A.}\ \bibnamefont {Ritchie}},\ and\ \bibinfo {author}
  {\bibfnamefont {A.~J.}\ \bibnamefont {Shields}},\ }\bibfield  {title}
  {\bibinfo {title} {A quantum light-emitting diode for the standard telecom
  window around 1,550 nm},\ }\href {https://doi.org/10.1038/s41467-018-03251-7}
  {\bibfield  {journal} {\bibinfo  {journal} {Nature communications}\ }\textbf
  {\bibinfo {volume} {9}},\ \bibinfo {pages} {862} (\bibinfo {year}
  {2018})}\BibitemShut {NoStop}%
\bibitem [{\citenamefont {Esposito}\ \emph {et~al.}(2017)\citenamefont
  {Esposito}, \citenamefont {Bietti}, \citenamefont {Fedorov}, \citenamefont
  {N{\"o}tzel},\ and\ \citenamefont {Sanguinetti}}]{Esposito.2017}%
  \BibitemOpen
  \bibfield  {author} {\bibinfo {author} {\bibfnamefont {L.}~\bibnamefont
  {Esposito}}, \bibinfo {author} {\bibfnamefont {S.}~\bibnamefont {Bietti}},
  \bibinfo {author} {\bibfnamefont {A.}~\bibnamefont {Fedorov}}, \bibinfo
  {author} {\bibfnamefont {R.}~\bibnamefont {N{\"o}tzel}},\ and\ \bibinfo
  {author} {\bibfnamefont {S.}~\bibnamefont {Sanguinetti}},\ }\bibfield
  {title} {\bibinfo {title} {Ehrlich-schw{\"o}bel effect on the growth dynamics
  of {G}a{A}s(111){A} surfaces},\ }\bibfield  {journal} {\bibinfo  {journal}
  {Physical Review Materials}\ }\textbf {\bibinfo {volume} {1}},\ \href
  {https://doi.org/10.1103/PhysRevMaterials.1.024602}
  {10.1103/PhysRevMaterials.1.024602} (\bibinfo {year} {2017})\BibitemShut
  {NoStop}%
\bibitem [{\citenamefont {Herzog}\ \emph {et~al.}(2012)\citenamefont {Herzog},
  \citenamefont {Bichler}, \citenamefont {Koblm{\"u}ller}, \citenamefont
  {Prabhu-Gaunkar}, \citenamefont {Zhou},\ and\ \citenamefont
  {Grayson}}]{Herzog.2012}%
  \BibitemOpen
  \bibfield  {author} {\bibinfo {author} {\bibfnamefont {F.}~\bibnamefont
  {Herzog}}, \bibinfo {author} {\bibfnamefont {M.}~\bibnamefont {Bichler}},
  \bibinfo {author} {\bibfnamefont {G.}~\bibnamefont {Koblm{\"u}ller}},
  \bibinfo {author} {\bibfnamefont {S.}~\bibnamefont {Prabhu-Gaunkar}},
  \bibinfo {author} {\bibfnamefont {W.}~\bibnamefont {Zhou}},\ and\ \bibinfo
  {author} {\bibfnamefont {M.}~\bibnamefont {Grayson}},\ }\bibfield  {title}
  {\bibinfo {title} {Optimization of {A}l{A}s/{A}l{G}a{A}s quantum well
  heterostructures on on-axis and misoriented {G}a{A}s(111){B}},\ }\href
  {https://doi.org/10.1063/1.4711783} {\bibfield  {journal} {\bibinfo
  {journal} {Applied Physics Letters}\ }\textbf {\bibinfo {volume} {100}},\
  \bibinfo {pages} {192106} (\bibinfo {year} {2012})}\BibitemShut {NoStop}%
\bibitem [{\citenamefont {Tuktamyshev}\ \emph {et~al.}(2019)\citenamefont
  {Tuktamyshev}, \citenamefont {Fedorov}, \citenamefont {Bietti}, \citenamefont
  {Tsukamoto},\ and\ \citenamefont {Sanguinetti}}]{Tuktamyshev.2019}%
  \BibitemOpen
  \bibfield  {author} {\bibinfo {author} {\bibfnamefont {A.}~\bibnamefont
  {Tuktamyshev}}, \bibinfo {author} {\bibfnamefont {A.}~\bibnamefont
  {Fedorov}}, \bibinfo {author} {\bibfnamefont {S.}~\bibnamefont {Bietti}},
  \bibinfo {author} {\bibfnamefont {S.}~\bibnamefont {Tsukamoto}},\ and\
  \bibinfo {author} {\bibfnamefont {S.}~\bibnamefont {Sanguinetti}},\
  }\bibfield  {title} {\bibinfo {title} {Temperature activated dimensionality
  crossover in the nucleation of quantum dots by droplet epitaxy on
  {G}a{A}s(111){A} vicinal substrates},\ }\href
  {https://doi.org/10.1038/s41598-019-51161-5} {\bibfield  {journal} {\bibinfo
  {journal} {Scientific Reports}\ }\textbf {\bibinfo {volume} {9}},\ \bibinfo
  {pages} {14520} (\bibinfo {year} {2019})}\BibitemShut {NoStop}%
\bibitem [{Note1()}]{Note1}%
  \BibitemOpen
  Note1,\ \href@noop {} {}\bibinfo {note} {See Supplemental Material at [URL
  will be inserted by publisher] for further details on the sample growth and
  morphology.}\BibitemShut {Stop}%
\bibitem [{\citenamefont {Young}\ \emph {et~al.}(2005)\citenamefont {Young},
  \citenamefont {Stevenson}, \citenamefont {Shields}, \citenamefont {Atkinson},
  \citenamefont {Cooper}, \citenamefont {Ritchie}, \citenamefont {Groom},
  \citenamefont {Tartakovskii},\ and\ \citenamefont {Skolnick}}]{Young.2005}%
  \BibitemOpen
  \bibfield  {author} {\bibinfo {author} {\bibfnamefont {R.~J.}\ \bibnamefont
  {Young}}, \bibinfo {author} {\bibfnamefont {R.~M.}\ \bibnamefont
  {Stevenson}}, \bibinfo {author} {\bibfnamefont {A.~J.}\ \bibnamefont
  {Shields}}, \bibinfo {author} {\bibfnamefont {P.}~\bibnamefont {Atkinson}},
  \bibinfo {author} {\bibfnamefont {K.}~\bibnamefont {Cooper}}, \bibinfo
  {author} {\bibfnamefont {D.~A.}\ \bibnamefont {Ritchie}}, \bibinfo {author}
  {\bibfnamefont {K.~M.}\ \bibnamefont {Groom}}, \bibinfo {author}
  {\bibfnamefont {A.~I.}\ \bibnamefont {Tartakovskii}},\ and\ \bibinfo {author}
  {\bibfnamefont {M.~S.}\ \bibnamefont {Skolnick}},\ }\bibfield  {title}
  {\bibinfo {title} {Inversion of exciton level splitting in quantum dots},\
  }\bibfield  {journal} {\bibinfo  {journal} {Physical Review B}\ }\textbf
  {\bibinfo {volume} {72}},\ \href {https://doi.org/10.1103/PhysRevB.72.113305}
  {10.1103/PhysRevB.72.113305} (\bibinfo {year} {2005})\BibitemShut {NoStop}%
\bibitem [{Note2()}]{Note2}%
  \BibitemOpen
  Note2,\ \href@noop {} {}\bibinfo {note} {See Supplemental Material at [URL
  will be inserted by publisher] for further details on the numerical
  methods.}\BibitemShut {Stop}%
\bibitem [{\citenamefont {Keating}(1966)}]{Keating.1966}%
  \BibitemOpen
  \bibfield  {author} {\bibinfo {author} {\bibfnamefont {P.~N.}\ \bibnamefont
  {Keating}},\ }\bibfield  {title} {\bibinfo {title} {Effect of invariance
  requirements on the elastic strain energy of crystals with application to the
  diamond structure},\ }\href {https://doi.org/10.1103/PhysRev.145.637}
  {\bibfield  {journal} {\bibinfo  {journal} {Physical Review}\ }\textbf
  {\bibinfo {volume} {145}},\ \bibinfo {pages} {637} (\bibinfo {year}
  {1966})}\BibitemShut {NoStop}%
\bibitem [{\citenamefont {Williamson}\ \emph {et~al.}(2000)\citenamefont
  {Williamson}, \citenamefont {Wang},\ and\ \citenamefont
  {Zunger}}]{Williamson.2000}%
  \BibitemOpen
  \bibfield  {author} {\bibinfo {author} {\bibfnamefont {A.~J.}\ \bibnamefont
  {Williamson}}, \bibinfo {author} {\bibfnamefont {L.~W.}\ \bibnamefont
  {Wang}},\ and\ \bibinfo {author} {\bibfnamefont {A.}~\bibnamefont {Zunger}},\
  }\bibfield  {title} {\bibinfo {title} {Theoretical interpretation of the
  experimental electronic structure of lens-shaped self-assembled
  {I}n{A}s/{G}a{A}s quantum dots},\ }\href
  {https://doi.org/10.1103/PhysRevB.62.12963} {\bibfield  {journal} {\bibinfo
  {journal} {Physical review. B, Condensed matter}\ }\textbf {\bibinfo {volume}
  {62}},\ \bibinfo {pages} {12963} (\bibinfo {year} {2000})}\BibitemShut
  {NoStop}%
\bibitem [{\citenamefont {Bester}(2009)}]{Bester.2009}%
  \BibitemOpen
  \bibfield  {author} {\bibinfo {author} {\bibfnamefont {G.}~\bibnamefont
  {Bester}},\ }\bibfield  {title} {\bibinfo {title} {Electronic excitations in
  nanostructures: an empirical pseudopotential based approach},\ }\href
  {https://doi.org/10.1088/0953-8984/21/2/023202} {\bibfield  {journal}
  {\bibinfo  {journal} {Journal of physics. Condensed matter : an Institute of
  Physics journal}\ }\textbf {\bibinfo {volume} {21}},\ \bibinfo {pages}
  {023202} (\bibinfo {year} {2009})}\BibitemShut {NoStop}%
\bibitem [{\citenamefont {Wang}\ and\ \citenamefont
  {Zunger}(1999)}]{Wang.1999}%
  \BibitemOpen
  \bibfield  {author} {\bibinfo {author} {\bibfnamefont {L.-W.}\ \bibnamefont
  {Wang}}\ and\ \bibinfo {author} {\bibfnamefont {A.}~\bibnamefont {Zunger}},\
  }\bibfield  {title} {\bibinfo {title} {Linear combination of bulk bands
  method for large-scale electronic structure calculations on strained
  nanostructures},\ }\href {https://doi.org/10.1103/PhysRevB.59.15806}
  {\bibfield  {journal} {\bibinfo  {journal} {Physical review. B, Condensed
  matter}\ }\textbf {\bibinfo {volume} {59}},\ \bibinfo {pages} {15806}
  (\bibinfo {year} {1999})}\BibitemShut {NoStop}%
\bibitem [{\citenamefont {Franceschetti}\ \emph {et~al.}(1999)\citenamefont
  {Franceschetti}, \citenamefont {Fu}, \citenamefont {Wang},\ and\
  \citenamefont {Zunger}}]{Franceschetti.1999}%
  \BibitemOpen
  \bibfield  {author} {\bibinfo {author} {\bibfnamefont {A.}~\bibnamefont
  {Franceschetti}}, \bibinfo {author} {\bibfnamefont {H.}~\bibnamefont {Fu}},
  \bibinfo {author} {\bibfnamefont {L.~W.}\ \bibnamefont {Wang}},\ and\
  \bibinfo {author} {\bibfnamefont {A.}~\bibnamefont {Zunger}},\ }\bibfield
  {title} {\bibinfo {title} {Many-body pseudopotential theory of excitons in
  {I}n{P} and {C}d{S}e quantum dots},\ }\href
  {https://doi.org/10.1103/PhysRevB.60.1819} {\bibfield  {journal} {\bibinfo
  {journal} {Physical review. B, Condensed matter}\ }\textbf {\bibinfo {volume}
  {60}},\ \bibinfo {pages} {1819} (\bibinfo {year} {1999})}\BibitemShut
  {NoStop}%
\bibitem [{\citenamefont {Resta}(1977)}]{Resta.1977}%
  \BibitemOpen
  \bibfield  {author} {\bibinfo {author} {\bibfnamefont {R.}~\bibnamefont
  {Resta}},\ }\bibfield  {title} {\bibinfo {title} {Thomas-fermi dielectric
  screening in semiconductors},\ }\href
  {https://doi.org/10.1103/PhysRevB.16.2717} {\bibfield  {journal} {\bibinfo
  {journal} {Physical Review B}\ }\textbf {\bibinfo {volume} {16}},\ \bibinfo
  {pages} {2717} (\bibinfo {year} {1977})}\BibitemShut {NoStop}%
\bibitem [{\citenamefont {Narvaez}\ \emph {et~al.}(2005)\citenamefont
  {Narvaez}, \citenamefont {Bester},\ and\ \citenamefont
  {Zunger}}]{Narvaez.2005}%
  \BibitemOpen
  \bibfield  {author} {\bibinfo {author} {\bibfnamefont {G.~A.}\ \bibnamefont
  {Narvaez}}, \bibinfo {author} {\bibfnamefont {G.}~\bibnamefont {Bester}},\
  and\ \bibinfo {author} {\bibfnamefont {A.}~\bibnamefont {Zunger}},\
  }\bibfield  {title} {\bibinfo {title} {Excitons, biexcitons, and trions in
  self-assembled ({I}n,{G}a){A}s/{G}a{A}s quantum dots: Recombination energies,
  polarization, and radiative lifetimes versus dot height},\ }\href
  {https://doi.org/10.1103/PhysRevB.72.245318} {\bibfield  {journal} {\bibinfo
  {journal} {Physical Review B}\ }\textbf {\bibinfo {volume} {72}},\ \bibinfo
  {pages} {245318} (\bibinfo {year} {2005})}\BibitemShut {NoStop}%
\bibitem [{\citenamefont {Sheng}(2006)}]{Sheng.2006}%
  \BibitemOpen
  \bibfield  {author} {\bibinfo {author} {\bibfnamefont {W.}~\bibnamefont
  {Sheng}},\ }\bibfield  {title} {\bibinfo {title} {Origins of optical
  anisotropy in artificial atoms},\ }\href {https://doi.org/10.1063/1.2370871}
  {\bibfield  {journal} {\bibinfo  {journal} {Applied Physics Letters}\
  }\textbf {\bibinfo {volume} {89}},\ \bibinfo {pages} {173129} (\bibinfo
  {year} {2006})}\BibitemShut {NoStop}%
\bibitem [{\citenamefont {Tonin}\ \emph {et~al.}(2012)\citenamefont {Tonin},
  \citenamefont {Hostein}, \citenamefont {Voliotis}, \citenamefont {Grousson},
  \citenamefont {Lemaitre},\ and\ \citenamefont {Martinez}}]{Tonin.2012}%
  \BibitemOpen
  \bibfield  {author} {\bibinfo {author} {\bibfnamefont {C.}~\bibnamefont
  {Tonin}}, \bibinfo {author} {\bibfnamefont {R.}~\bibnamefont {Hostein}},
  \bibinfo {author} {\bibfnamefont {V.}~\bibnamefont {Voliotis}}, \bibinfo
  {author} {\bibfnamefont {R.}~\bibnamefont {Grousson}}, \bibinfo {author}
  {\bibfnamefont {A.}~\bibnamefont {Lemaitre}},\ and\ \bibinfo {author}
  {\bibfnamefont {A.}~\bibnamefont {Martinez}},\ }\bibfield  {title} {\bibinfo
  {title} {Polarization properties of excitonic qubits in single self-assembled
  quantum dots},\ }\href {https://doi.org/10.1103/PhysRevB.85.155303}
  {\bibfield  {journal} {\bibinfo  {journal} {Physical Review B}\ }\textbf
  {\bibinfo {volume} {85}},\ \bibinfo {pages} {3050} (\bibinfo {year}
  {2012})}\BibitemShut {NoStop}%
\bibitem [{\citenamefont {Gao}\ \emph {et~al.}(2022)\citenamefont {Gao},
  \citenamefont {Rickert}, \citenamefont {Urban}, \citenamefont {Große},
  \citenamefont {Srocka}, \citenamefont {Rodt}, \citenamefont {Musial},
  \citenamefont {Zolnacz}, \citenamefont {Mergo}, \citenamefont {Dybka},
  \citenamefont {Urbańczyk}, \citenamefont {Sek}, \citenamefont {Burger},
  \citenamefont {Reitzenstein},\ and\ \citenamefont {Heindel}}]{Gao.2022}%
  \BibitemOpen
  \bibfield  {author} {\bibinfo {author} {\bibfnamefont {T.}~\bibnamefont
  {Gao}}, \bibinfo {author} {\bibfnamefont {L.}~\bibnamefont {Rickert}},
  \bibinfo {author} {\bibfnamefont {F.}~\bibnamefont {Urban}}, \bibinfo
  {author} {\bibfnamefont {J.}~\bibnamefont {Große}}, \bibinfo {author}
  {\bibfnamefont {N.}~\bibnamefont {Srocka}}, \bibinfo {author} {\bibfnamefont
  {S.}~\bibnamefont {Rodt}}, \bibinfo {author} {\bibfnamefont {A.}~\bibnamefont
  {Musial}}, \bibinfo {author} {\bibfnamefont {K.}~\bibnamefont {Zolnacz}},
  \bibinfo {author} {\bibfnamefont {P.}~\bibnamefont {Mergo}}, \bibinfo
  {author} {\bibfnamefont {K.}~\bibnamefont {Dybka}}, \bibinfo {author}
  {\bibfnamefont {W.}~\bibnamefont {Urbańczyk}}, \bibinfo {author}
  {\bibfnamefont {G.}~\bibnamefont {Sek}}, \bibinfo {author} {\bibfnamefont
  {S.}~\bibnamefont {Burger}}, \bibinfo {author} {\bibfnamefont
  {S.}~\bibnamefont {Reitzenstein}},\ and\ \bibinfo {author} {\bibfnamefont
  {T.}~\bibnamefont {Heindel}},\ }\bibfield  {title} {\bibinfo {title} {A
  quantum key distribution testbed using a plug\&play telecom-wavelength
  single-photon source},\ }\href {https://doi.org/10.1063/5.0070966} {\bibfield
   {journal} {\bibinfo  {journal} {Applied Physics Reviews}\ ,\ \bibinfo
  {pages} {011412}} (\bibinfo {year} {2022})}\BibitemShut {NoStop}%
\bibitem [{\citenamefont {Blokhin}\ \emph {et~al.}(2021)\citenamefont
  {Blokhin}, \citenamefont {Bobrov}, \citenamefont {Maleev}, \citenamefont
  {Donges}, \citenamefont {Bremer}, \citenamefont {Blokhin}, \citenamefont
  {Vasil'ev}, \citenamefont {Kuzmenkov}, \citenamefont {Kolodeznyi},
  \citenamefont {Shchukin}, \citenamefont {Ledentsov}, \citenamefont
  {Reitzenstein},\ and\ \citenamefont {Ustinov}}]{Blokhin.2021}%
  \BibitemOpen
  \bibfield  {author} {\bibinfo {author} {\bibfnamefont {S.~A.}\ \bibnamefont
  {Blokhin}}, \bibinfo {author} {\bibfnamefont {M.~A.}\ \bibnamefont {Bobrov}},
  \bibinfo {author} {\bibfnamefont {N.~A.}\ \bibnamefont {Maleev}}, \bibinfo
  {author} {\bibfnamefont {J.~N.}\ \bibnamefont {Donges}}, \bibinfo {author}
  {\bibfnamefont {L.}~\bibnamefont {Bremer}}, \bibinfo {author} {\bibfnamefont
  {A.~A.}\ \bibnamefont {Blokhin}}, \bibinfo {author} {\bibfnamefont {A.~P.}\
  \bibnamefont {Vasil'ev}}, \bibinfo {author} {\bibfnamefont {A.~G.}\
  \bibnamefont {Kuzmenkov}}, \bibinfo {author} {\bibfnamefont {E.~S.}\
  \bibnamefont {Kolodeznyi}}, \bibinfo {author} {\bibfnamefont {V.~A.}\
  \bibnamefont {Shchukin}}, \bibinfo {author} {\bibfnamefont {N.~N.}\
  \bibnamefont {Ledentsov}}, \bibinfo {author} {\bibfnamefont {S.}~\bibnamefont
  {Reitzenstein}},\ and\ \bibinfo {author} {\bibfnamefont {V.~M.}\ \bibnamefont
  {Ustinov}},\ }\bibfield  {title} {\bibinfo {title} {{Design optimization for
  bright electrically-driven quantum dot single-photon sources emitting in
  telecom O-band}},\ }\href {https://doi.org/10.1364/OE.415979} {\bibfield
  {journal} {\bibinfo  {journal} {{Optics express}}\ }\textbf {\bibinfo
  {volume} {29}},\ \bibinfo {pages} {6582} (\bibinfo {year}
  {2021})}\BibitemShut {NoStop}%
\bibitem [{\citenamefont {Bremer}\ \emph {et~al.}(2022)\citenamefont {Bremer},
  \citenamefont {Jimenez}, \citenamefont {Thiele}, \citenamefont {Weber},
  \citenamefont {Huber}, \citenamefont {Rodt}, \citenamefont {Herkommer},
  \citenamefont {Burger}, \citenamefont {H{\"o}fling}, \citenamefont
  {Giessen},\ and\ \citenamefont {Reitzenstein}}]{Bremer.2022}%
  \BibitemOpen
  \bibfield  {author} {\bibinfo {author} {\bibfnamefont {L.}~\bibnamefont
  {Bremer}}, \bibinfo {author} {\bibfnamefont {C.}~\bibnamefont {Jimenez}},
  \bibinfo {author} {\bibfnamefont {S.}~\bibnamefont {Thiele}}, \bibinfo
  {author} {\bibfnamefont {K.}~\bibnamefont {Weber}}, \bibinfo {author}
  {\bibfnamefont {T.}~\bibnamefont {Huber}}, \bibinfo {author} {\bibfnamefont
  {S.}~\bibnamefont {Rodt}}, \bibinfo {author} {\bibfnamefont {A.}~\bibnamefont
  {Herkommer}}, \bibinfo {author} {\bibfnamefont {S.}~\bibnamefont {Burger}},
  \bibinfo {author} {\bibfnamefont {S.}~\bibnamefont {H{\"o}fling}}, \bibinfo
  {author} {\bibfnamefont {H.}~\bibnamefont {Giessen}},\ and\ \bibinfo {author}
  {\bibfnamefont {S.}~\bibnamefont {Reitzenstein}},\ }\bibfield  {title}
  {\bibinfo {title} {{Numerical optimization of single-mode fiber-coupled
  single-photon sources based on semiconductor quantum dots}},\ }\href
  {https://doi.org/10.1364/OE.456777} {\bibfield  {journal} {\bibinfo
  {journal} {{Optics express}}\ }\textbf {\bibinfo {volume} {30}},\ \bibinfo
  {pages} {15913} (\bibinfo {year} {2022})}\BibitemShut {NoStop}%
\bibitem [{\citenamefont {Rickert}\ \emph {et~al.}(2019)\citenamefont
  {Rickert}, \citenamefont {Kupko}, \citenamefont {Rodt}, \citenamefont
  {Reitzenstein},\ and\ \citenamefont {Heindel}}]{Rickert.2019}%
  \BibitemOpen
  \bibfield  {author} {\bibinfo {author} {\bibfnamefont {L.}~\bibnamefont
  {Rickert}}, \bibinfo {author} {\bibfnamefont {T.}~\bibnamefont {Kupko}},
  \bibinfo {author} {\bibfnamefont {S.}~\bibnamefont {Rodt}}, \bibinfo {author}
  {\bibfnamefont {S.}~\bibnamefont {Reitzenstein}},\ and\ \bibinfo {author}
  {\bibfnamefont {T.}~\bibnamefont {Heindel}},\ }\bibfield  {title} {\bibinfo
  {title} {{Optimized designs for telecom-wavelength quantum light sources
  based on hybrid circular Bragg gratings}},\ }\href
  {https://doi.org/10.1364/OE.27.036824} {\bibfield  {journal} {\bibinfo
  {journal} {{Optics express}}\ }\textbf {\bibinfo {volume} {27}},\ \bibinfo
  {pages} {36824} (\bibinfo {year} {2019})}\BibitemShut {NoStop}%
\bibitem [{\citenamefont {Barbiero}\ \emph
  {et~al.}(2022{\natexlab{a}})\citenamefont {Barbiero}, \citenamefont {Huwer},
  \citenamefont {Skiba-Szymanska}, \citenamefont {M{\"u}ller}, \citenamefont
  {Stevenson},\ and\ \citenamefont {Shields}}]{Barbiero.2022}%
  \BibitemOpen
  \bibfield  {author} {\bibinfo {author} {\bibfnamefont {A.}~\bibnamefont
  {Barbiero}}, \bibinfo {author} {\bibfnamefont {J.}~\bibnamefont {Huwer}},
  \bibinfo {author} {\bibfnamefont {J.}~\bibnamefont {Skiba-Szymanska}},
  \bibinfo {author} {\bibfnamefont {T.}~\bibnamefont {M{\"u}ller}}, \bibinfo
  {author} {\bibfnamefont {R.~M.}\ \bibnamefont {Stevenson}},\ and\ \bibinfo
  {author} {\bibfnamefont {A.~J.}\ \bibnamefont {Shields}},\ }\bibfield
  {title} {\bibinfo {title} {{Design study for an efficient semiconductor
  quantum light source operating in the telecom C-band based on an
  electrically-driven circular Bragg grating}},\ }\href
  {https://doi.org/10.1364/OE.452328} {\bibfield  {journal} {\bibinfo
  {journal} {{Optics express}}\ }\textbf {\bibinfo {volume} {30}},\ \bibinfo
  {pages} {10919} (\bibinfo {year} {2022}{\natexlab{a}})}\BibitemShut {NoStop}%
\bibitem [{\citenamefont {Barbiero}\ \emph
  {et~al.}(2022{\natexlab{b}})\citenamefont {Barbiero}, \citenamefont {Huwer},
  \citenamefont {Skiba-Szymanska}, \citenamefont {Ellis}, \citenamefont
  {Stevenson}, \citenamefont {M{\"u}ller}, \citenamefont {Shooter},
  \citenamefont {Goff}, \citenamefont {Ritchie},\ and\ \citenamefont
  {Shields}}]{Barbiero.2022b}%
  \BibitemOpen
  \bibfield  {author} {\bibinfo {author} {\bibfnamefont {A.}~\bibnamefont
  {Barbiero}}, \bibinfo {author} {\bibfnamefont {J.}~\bibnamefont {Huwer}},
  \bibinfo {author} {\bibfnamefont {J.}~\bibnamefont {Skiba-Szymanska}},
  \bibinfo {author} {\bibfnamefont {D.~J.~P.}\ \bibnamefont {Ellis}}, \bibinfo
  {author} {\bibfnamefont {R.~M.}\ \bibnamefont {Stevenson}}, \bibinfo {author}
  {\bibfnamefont {T.}~\bibnamefont {M{\"u}ller}}, \bibinfo {author}
  {\bibfnamefont {G.}~\bibnamefont {Shooter}}, \bibinfo {author} {\bibfnamefont
  {L.~E.}\ \bibnamefont {Goff}}, \bibinfo {author} {\bibfnamefont {D.~A.}\
  \bibnamefont {Ritchie}},\ and\ \bibinfo {author} {\bibfnamefont {A.~J.}\
  \bibnamefont {Shields}},\ }\bibfield  {title} {\bibinfo {title}
  {High-performance single-photon sources at telecom wavelength based on
  broadband hybrid circular bragg gratings},\ }\bibfield  {journal} {\bibinfo
  {journal} {ACS Photonics}\ }\href
  {https://doi.org/10.1021/acsphotonics.2c00810} {10.1021/acsphotonics.2c00810}
  (\bibinfo {year} {2022}{\natexlab{b}})\BibitemShut {NoStop}%
\end{thebibliography}%

\end{document}